\documentclass[proof]{WileyASNA-v1}

\articletype{Original article}%

\received{xx xxxxx xxxx}
\revised{xx xxxxx xxxx}
\accepted{xx xxxxx xxxx}

\begin{document}

\title{Review of light curves of novae in the modified scales. I. Recurrent novae}

\author[1]{Rosenbush A.*}




\address[1]{\orgname{Main Astronomical Observatory of the National Academy of Sciences of Ukraine}, \orgaddress{\state{27 Akademika Zabolotnoho St., 03143 Kyiv}, \country{Ukraine}}}



\corres{*\email{aeros@mao.kiev.ua}}


\abstract{We reviewed the light curves of 10 recurrent novae in the modified scales of the "logarithmic of time, amplitude of outburst". A unified rule was established to normalize the time scale. 

Confirmation of the known result: in a series of recurrent outbursts, the light curve of nova retains its shape, was realized by distinguishing three groups amongst these 10 novae. The CI Aql (+V2487 Oph) and T Pyx (+ IM Nor) groups contain by one additional member. The V745 Sco group includes V394 CrA and V3890 Sgr. Three RNs: T CrB, RS Oph and U Sco, have its own uniqueness. 

After comparing the light curves of galactic novae with the light curves of some novae of the Large and Small Magellanic Clouds (LMC and SMC), we selected several candidates for recurrent novae in addition to 3 known recurrent novae in these galaxies. Their belonging to a certain group or the likeness of a specific galactic nova was specified. Assuming that the physical characteristics of the novae in the group coincide, we estimated the absolute magnitudes of the galactic novae: CI Aql and V2847 Oph have absolute magnitudes M$_{V}$=-7.5$^{m}$, IM Nor and T Pyx - M$_{V}$=-7.2$^{m}$, V3890 Sgr - M$_{V}$=-7.9$^{m}$, U Sco - M$_{I}$=-9.0$^{m}$.  

A comparison of the summarized light curves of recurrent novae with the light curves of some low-amplitude classical novae allowed us amongst the last to distinguish the list of 26 candidates in recurrent novae. Three of them are known from lists of other authors. An interesting example is the V1017 Sgr, which erupted in 1919 as a nova of the T Pyx type, and which often also shows less bright outbursts classified as dwarf nova outbursts.}


\jnlcitation{\cname{%
\author{Rosenbush A.}} (\cyear{2020}), 
\ctitle{Review of light curves of novae in the modified scales. I. Recurrent novae}, \cjournal{Astron. Nachr.}, \cvol{20--;00:1--6}.}


\maketitle


\section{Introduction}\label{sec1}

In certain binary systems, under certain conditions, a short-term surface source of thermonuclear energy may arise \citep{Starrfield2016}. As a result, the phenomenon of a "new" star or a nova is observed. By the outburst amplitude of more than 9-11$^{m}$, the nova is classified as classical nova (CN). 

A small number of novae after some time intervals recurrent outbursts of a less high amplitude of brightness changes. These novae are known as recurrent novae (RNs). 

The list of galactic recurrent novae contains 10 objects: CI Aql, V394 CrA, T CrB, IM Nor, RS Oph, V2487 Oph, T Pyx, V3890 Sgr, U Sco, and V745 Sco. The duration of a outburst from the beginning to the return of the star system to a quiet state takes from about 70 days (U Sco, V3890 Sgr) to 2-3 years (IM Nor, T Pyx) against 3-10 or more years in classical novae. Outbursts of recurrent novae recur over certain long periods of time up to hundred years. There is an opinion that classical novae also experience recurrent outbursts, but separated by intervals of hundreds to thousands of years. A comprehensive study of all known outbursts of all known recurrent novae was performed by \cite{Schaefer2010a}. 

The list of classical novae is replenished every year with new objects. To date, the number of known classical novae of our Galaxy is approaching 450. The issue of systematizing these objects has been resolved with the accumulation of information and still remains relevant. 

The development of flares was traditionally presented as the dependence of star brightness on time. General similar phases of the development of brightness and spectrum in the outburst were distinguished: initial and final rises, principal maximum, early decline, transition phase, final decline, and post-nova \citep{McLaughlin1942, Payne-Gaposchkin1957}. The duration of the phases was different, they occurred at different brightness and at different times after the outburst, therefore there was added a characteristic  of the fast or slow nova with gradation into 5 classes, which became the main classification scheme (see, for example, \citep{Payne-Gaposchkin1957}). The key parameter in this scheme is the parameter t$_{m}$: the time of brightness decline in days by 2$^{m}$ or 3$^{m}$ after the principal maximum. Within each speed class, there are noticeable differences in the behavior of the nova brightness: the light curve at the transition phase of the outburst can be, for example, either smooth, or with oscillations, or with a temporary deep decline of brightness. The temporal decline of the brightness at the transition phase was the most contrasting and recognizable detail of the light curve, which gave the ground in order to similar novae to call ''DQ Her type'', after the first such well-studied Nova Herculis of 1934 (see, for example, \citep{Payne-Gaposchkin1957}). 

The presence of similar details of the light curves of various novae at certain phases of the outburst stimulated the search for classification schemes based on the shape of the light curve. \cite{McLaughlin1960} pointed out that a corresponding compression of the timeline would make it possible to coincide the light curves for some novae. He recognized the attempt to unify the light curves in units of the tm parameter as unsuccessful: slow novae reached late phases earlier than fast ones. Novae of the DQ Her type were in the detailed scheme of \cite{Duerbeck1981, Duerbeck1987a}, they are also present in the scheme of \cite{Strope2010}. Whereas in the first scheme, recurrent novae were mentioned separately from classical novae, then in the recently proposed scheme of \cite{Strope2010} recurrent novae were distributed between distinguished types of classical novae with common details of light curves. 

These classification schemes are based on the light curve in the traditional linear representation of the time scale. Sometimes, a logarithmic representation of the time scale was used to provide a more visual representation of some features of the brightness change of novae. This representation is used, in particular, when extrapolating the light curve to estimate the brightness of a nova at an outburst maximum. For example, the logarithmic representation was used to derive a very important result: the absolute brightness of the novae of different speed types on the 15th day after the outburst maximum is approximately the same \citep{Buscombe1955}. The logarithmic representation of the light curve has been partially used in the study of \cite{Strope2010}. A similar representation of the abscissa scale is used by \cite{Hachisu2018} when modelling outbursts of novae; note that their set of novae includes both very fast novae and well-known recurrent novae, including those that will be proposed as candidates for recurrent novae in our current study. 

The existing systematization schemes of novae by the shape of the light curve did not give an answer to the question of the relationship between the shape of the light curve and the physical properties of the novae. \cite{Strope2010} noted “the primary reason why none of these classification schemes has been usefully adopted is that the classes are so broad that greatly different physical settings are lumped together (heaped together) and they have not been found to correlate with anything else”. 

The aim of our review is also the question of finding the answer to the question of whether it is possible to systematize novae based on the shape of the light curve. But for this we have changed the approach to the presentation of light curves \citep{Rozenbush1996, Rosenbush1999a, Rosenbush1999b}. 

In our initial study of the feasibility of constructing a classification scheme for novae by the shape of the light curve \citep{Rosenbush1999a, Rosenbush1999b}, we proposed moving from a timeline to a scale of the radius of the shell ejected during the outburst \cite{Rozenbush1996}. This made it possible to somewhat bring close together, but not coincide, the same phases of the light curve of novae on the abscissa scale. For a compact representation of the prolonged process, we moved on to the logarithmic representation of the abscissa scale. [In the catalog of \cite{Strope2010}, the logarithmic representation of the timeline — abscissa - is recognized as “the only way that the late-time data can be usefully displayed.”] The positive result of the last transformation is a more compact and easy-to-view presentation of the complete the light curve of a nova outburst with well noticeable fast light changes near the light maximum and also very noticeable very slow changes in the final phases, which greatly facilitated the comparison of light curves of different novae. It was proposed to normalize the radius scale to the radius of the nova at the principal maximum r$_{0}$, for which a value of 1.4$\times$10$^{13}$ cm or log(r$_{0}$)$\approx$13.1  for all novae was taken \citep{Rosenbush1999d}. 

Another modification is related to the ordinate scale: from the current brightness of the object, we moved to the amplitude of the outburst, i.e. the null-point of the ordinate scale was attributed to the brightness of the quiet state of a nova \citep{Rosenbush1999a, Rosenbush1999b}. 

The prerequisites for the practical application of this view in research of novae can be found, for example, in a theoretical study by \cite{Gordeladse1937}, where it was concluded that the temporal dependences of the luminosity of a gas cloud of various geometric shapes which  is expanding with moving away from the star had a power dependence with certain degrees. In a more recent study, \cite{Slavin1995} concluded that faster novae have randomly distributed clumps of matter in a spherically symmetric diffuse ejection, while slower novae have more structured ellipsoids with one or more enhanced emission rings. 

Our monitoring of photometric data of novae after an initial study \citep{Rosenbush1999a, Rosenbush1999b} confirmed the prospects of this approach in constructing a classification scheme for novae. There have been cases of almost identical light curves among classical novae. An attempt to apply it to low-amplitude objects encountered a wide variety of details of light curves with small ranges of light changes \citep{Rosenbush1999c}. The application of our approach to recurrent novae allowed us to suspect the recurrent nature of V2487 Oph for which the search of \cite{Pagnotta2009} in the photographic archives gave as result the discovery of the outburst of 1900.   

While the study of \cite{Rosenbush1999a, Rosenbush1999b} only draw attention to some general characteristics of the novae in one group, then as \cite{Schaefer2010a}, as a result of the comprehensive study of recurrent novae, made a number of general conclusions, including one that is very important for us: ''All the eruption light curves from a single RN are consistent with a single invariant template. That is, the light curves are always the same from eruption to eruption. This tells us that the eruption light curve depends on system parameters (like the white dwarf mass and the composition) that do not vary from eruption to eruption (like the mass loss rate)''. It is important that here is noted the stability of the light curve in outbursts and the dependence of the light curve from the parameters of the binary system in which these outbursts occur. 

This conclusion may mean that if we find two novae with identical light curves, then these objects form a group. 

The result obtained \cite{Schaefer2010a} on a linear time scale, we repeat in our modification of the light curves and as a result we will find similarities in the light curves of some recurrent novae, which will give us a basis for combining them into a group. 

This review of light curves of novae begins with a review of modified light curves of recurrent novae. We more fully use the sense of the word "recurrent", including this: recurrence in one property or parameter implies recurrence in other properties or parameters. 

Last doubts about the application of this approach were eliminated by the outburst of the unique classical Nova Scorpii 2007, V1280 Sco, which we will examine in detail in the second part of our study. The light curve of this nova showed a temporal decline in light of complex shape. This classical nova, along with our analysis of the recurrent novae presented below, convinced us that there was no need to shift along the abscissa axis the light curves when comparing the light curves of two novae. Small deviations can reflect real differences in the parameters of two novae of the same type or group, and large deviations should be considered as a sign of another type, or group, of novae. It is important to note that the initial use of the shift to coincide light curves in the studies of \cite{Rosenbush1999a,Rosenbush1999b} did not have a significant negative impact on the definition and composition of the main groups of novae in this study. For the abscissa scale, it was decided to keep the use of the radius of the ejected shell, since this gives an estimate of the size of the shell ejected during the outburst, especially since at the final phase of the outburst the radiation of a nova in the visual region of the spectrum is mainly determined by the radiation of this expanding shell (for more details see, e.g., monograph of \cite{Vorontsov-Velyaminov1953}). Therefore, our first step will be to formulate a rule for unambiguous determining the instant of time for the maximum brightness in a outburst and this will be the only parameter that we will sometimes set when constructing a modified light curve.  

Such very active and large-scale phenomena as outbursts of novae are based on thermonuclear reactions that require certain conditions to be met (see, for example, the review of \cite{Starrfield2016}). And since the conditions in the sequence of outbursts for one recurrent nova will be the same, the energy released during recurrent outbursts, or the bolometric luminosity of a nova in a series of outbursts, will be constant or almost constant. This means that the light curve of a nova can ultimately be represented in units of absolute energy. 

The light curve is a change in the radiation of a star over time. The radiation of a nova, which is an exploded component of a binary star system with circumstellar and near-system structures, has several main components: 1 - the hot acceptor star itself, which erupted (first expanding and then returning to its normal state), 2 - the ejected shell (re-radiating energy flux the central active source), 3 - the accretion disk and other circumstellar and near-system formations, 4 - the second donor star. The contribution of each of these components to the observed process varies with time, and their ratio is different for different novae. 

Recurrent novae are considered as a separate type (see, for example, \citep{Duerbeck1981, Duerbeck1987a}). During the outburst of a recurrent nova the large-scale ejection of the shell does not occur. In the future, we will use the properties of members this novae type as one of the main arguments to justify the validity of our approach to the study of classical novae. The recurrence of outbursts in these recurrent objects occurs with the recurrence of many details of the light curve. The difference in light curves is partly due to differences in instrumental photometric systems; the noise band of a single light curve characterizes the accuracy of the data. The scatter of composed light curve of novae for a particular group may characterize possible changes in the outburst parameters. 

In the course of this study, it will be shown that (1) the light curves of the recurrent nova in the observed sequence of outbursts are identical in detail or practically coincide, that (2) some of the recurrent novae have similar light curves, that is, we have the case of two recurrent novae as twins. Some differences may arise due to difference in the spatial orientation of the erupted binary system relative to the observer \citep{Warner1987}. Similarly, proceeding to the study of classical novae, we can assume that if there is a certain group of novae with similar light curves, then in each group we are dealing with the same configuration of a binary system, up to the distribution of matter flowing from a stellar donor in the vicinity and on the surface of the acceptor. Identification of new members of a certain group in nearby galaxies will allow us to find out the absolute luminosity of such novae with minimal assumptions and errors. 

Eruptive stars of other types with a smaller outburst amplitude than classical and recurrent novae are not affected in this study. But dwarf novae such as WZ Sge and especially ''tremendous outburst amplitude dwarf novae'' or TOADs \citep{Howell1995, Kato2015} were taken into account in the preliminary classification: the amplitude and duration of the state of maximum brightness were comparable. 

The basis of this study is the observations contained in numerous publications and extensive publicly available databases of international associations of observers of variable stars (AAVSO, AFOEV, BAAVSS). The main source, AAVSO data \citep{Kafka2019}, was supplemented by OGLE sky survey data \citep{Kozlowski2013,Wyrzykowski2014}\footnote{http://ogle.astrouw.edu.pl/ogle4/transients/}, in particular in the galactic bulge region \citep{Mroz2015} and the Large and Small Magellanic clouds (LMC and SMC) \citep{Mroz2016a}, the specialized survey SMARTS \citep{Walter2012}, the ASAS-SN sky survey \citep{Kochanek2017, Shappee2014} and others. For recurrent novae, there is an extensive compilation of observations with a substantial addition of data for historical outbursts and a detailed discussion of individual stars \citep{Schaefer2010a}. In addition to the SIMBAD database, catalogues and monographs served as sources of the bibliography on objects: VSX AAVSO \citep{Kafka2019}, \cite{Payne-Gaposchkin1957, Samus2017, Duerbeck1987a, Strope2010}, \cite{Pietsch2010}\footnote{http://www.mpe.mpg.de/\_m31novae/opt/index.php}. 

Classification at this phase occurs through visual comparison with a set of templates of ''standard'' light curves.

\section{The procedure of constructing light curves of novae in modified scales}\label{sec2}

To estimate the radius of the shell ejected during a nova outburst, knowledge of the shell expansion velocity V${_{exp}}$ is necessary. But since this value is often not known and, most likely, it is not a constant value, the logarithmic scale of the shell radius was changed to the logarithmic scale of the time with a subsequent normalization. 

The radius of the ejected shell is generally represented by the expression 
 
log(r)=log[r${_0}$+(t-t${_0}$)$\times$V${_{exp}}$]=log[(t-t${_0}$)$\times$V${_{exp}}$]+
${ }$
+log[1+r${_0}$/(t-t${_0}$)/V${_{exp}}$], ${    }$        (1) 

where t is the current time, t${_0}$ is the moment of maximal brightness. 

We transform and simplify the expression (1)

log(r)=log(t-t${_0}$)+log(V${_{exp}}$)+$\Delta$(t).    ${    }$      (1a) 

The term $\Delta$(t) tends to zero with increasing of t, log(V${_{exp}}$) is assumed to be constant. 

For the practical use, we rewrite the expression (1a) in the form 

log(r)=log(t-t${_0}$)+C,    ${    }$       (2) 

where C$\approx$log(r${_0}$)=13.1, i.e. formally, this is the radius of the shell on the first day after the maximal brightness, when a separation of the main shell from the nova occur. Most likely, the value of the constant C varies within one or two orders of magnitude (the errors of the constant C, large on the first day, decrease with time, see below.) The time t and the radius of the shell r enter into expression (2) in days and centimeters, respectively. The parameter t${_0}$ will define as an integer; with fractions of the day we will operate only in the case of very fast outbursts. 

This simplification gives us an idea of the radius of the expanding shell, the radiation and processes of which forms a light curve, starting from the transition phase. 

In fact, the brightness of the nova is a function of several arguments: the temperature of the central source, the density of the substance in the expanding shell, etc. \citep{Hachisu2006}. If there is no the ejection of the envelope, that takes place in the case of recurrent novae, then the obtained light curve characterizes the process of the returning of a star to its normal state. The only circumstellar envelope is observed near the recurrent nova T Pyx \citep{Duerbeck1979} and it is proposed to associate it with a unique outburst of 1866 \citep{Schaefer2010b}.

It was proposed to normalize the scale of the current brightness of the nova to the brightness of the quiet state of the binary system, before or after the outburst \citep{Rosenbush1999a, Rosenbush1999b}. That is the ordinate scale has become the scale of outburst amplitude. This is closer to understanding that the development of an outburst depends on the energy released by the burning of a thermonuclear source. In final form, the modified scales for the light curves of recurrent novae have the form “time logarithm, outburst amplitude” \citep{Rosenbush1999a, Rosenbush1999b}. 

For the aims of our review, we will use the very useful properties of the logarithmic representation of the abscissa scale for the light curves of these eruptive stars:

1 - linearization of some sections of light curves;

2 - compact representation of very long-lasting events: stretching of fast changes of brightness in a state of maximum and compression of very slow changes of brightness at the end of an outburst;

3 - weak dependence of the shape of the light curve at the end of the outburst on errors in determining the moment of principal maximum outburst, which is important in the case of a missed maximum of outburst. 

\cite{Rosenbush1999d} discussed the effect of inaccuracies and uncertainties in determining certain parameters on the shape of the light curve in the modified scales. But within the same group of novae, many errors are counterbalanced by a uniform definition of the necessary parameters. In connection with the need for a uniform determination of the time moment t${_0}$ and for the convenience of further exposition, it seems to us necessary, in addition to the concepts of the phases of a schematic light curve of a nova \citep{McLaughlin1942, Payne-Gaposchkin1957}, to introduce the concept of “state of maximum brightness”. This state includes the phase of the final rise to the principal maximum and the principal maximum itself. After a state of maximum brightness, a steady trend is formed for the next phase - the phase of an early decline in brightness. For slow novae, for example RR Pic, this state of maximum brightness can last up to hundreds of days and include several equivalent local maximums of brightness. Therefore, for slow novae, the time moment t${_0}$ will be defined as the moment of transition from the fast initial brightness rise to the slower final rise or to the state of maximum brightness. For very fast novae the fixation of the moment t${_0}$ is unique. For our aims, when determining the moment t${_0}$, the accuracy of 10\% of the value of t-t${_0}$ is sufficient. For details of the light curve near the maximum, the duration of which is $\le$1${^d}$, this is an accuracy of 0.1${^d}$. If the state of maximum brightness has the duration of the order of 100${^d}$, then the error 10${^d}$ in t${_0}$ is allowable. 

Show this on the extreme cases that were occured in our review. In the modern era of monitoring observations, the maximum moment t${_0}$ for fast novae is often known with the precision of the order of a day or even fractions of a day. Seasonal gaps due to the conjuction of the object with the Sun reach a hundred of days. Outbursts of fast recurrent novae last for 50-100 days, therefore it is advisable to know t${_0}$ up to a fraction of a day: a day after the maximum the shift of this point of light curve along the abscissa axis because of the this error $\sigma$ can be $\Delta$log(t-t${_0}$)$\equiv$$\Delta$log(1+$\sigma$)$\equiv$$\Delta$log(1+0.1)$\le$0.04, but will decrease with the increase of t-t${_0}$. Thus, the stable shape of the light curve in the logarithmic time scale is formed already from the first day and after 10-15 days of the outburst we can talk about the invariable shape of the light curve. Therefore, for a recurrent outburst with a missed maximum, this later part of the light curve can reasonably be shifted along a logarithmic time scale to best match with a more complete light curve in another outburst. The light curve of a fast nova in the vicinity of the light curve maximum is possible to reconstruct by combining several outbursts by varying the moment t${_0}$ for each outburst with an accuracy of a fraction of a day, since the recurrence of the shape of the light curve after the maximum brightness mean that the shape is also preserved by the maximum brightness as well. In the process of classifying a specific object, the numerical value of t${_0}$ may decrease or increase. The restriction for the reduction of t${_0}$, i.e. (t-t${_0}$) will increase, it is determined by the appearance of an increasing deviation of the initial portion of the light curve to the right and up from a straight line approximating the subsequent initial decrease of the brightness, i.e. the light curve in the region of maximum show an ascent up and to the right. Accordingly, an increase in t${_0}$, i.e. (t-t${_0}$) will decrease, lead to a decrease in the slope of the initial part of the light curve and an increase in its length to the left. Thus, the procedure for choosing the moment t${_0}$ for fast novae can lead to a formal shift of the maximum of the light curve by log(t-t${_0}$)+C<13.1, but this will not lead to a noticeable shift of the subsequent part of the light curve and to the change in its shape. In the course of our research, this happened in several rare cases (about a dozen to three hundred or four hundreds novae that have passed our qualifications). This circumstance was noted as a drawback by \cite{Bode2016} by using a light curve with a logarithmic representation of the time scale in the study of LMCN 2009-02a. It is important that the determination of parameters within a specific group of fast or slow novae proceeds by the uniform procedure. 

It is also desirable to know the date of the last observation of a pre-nova, which will also indicate the second, permissible limits of variation of t${_0}$. 

When comparing the final segments of the light curves in the modified scales, it should also be taken into account that the formation of these segments occurs in the presence of an increasing contribution of the second component of the binary system and possibly other sources. This can be suspected if, starting from a level of 2.5${^m}$ above a quiet state, the light curve will show a slowdown in the rate of light decline. 

The third outburst of V3890 Sgr provided a unique opportunity to test our methodology and evaluate its errors in practice, since it occurred already after the results of the study were prepared for publication and the parameters of the two previous outbursts were consistent with each other according to the above. It is important that the outburst was detected by \cite{Pereira2019} at almost the maximum brightness of JD 2458723.37. Our estimate of the parameter t${_0}$ (Table 1) by -0.11${^d}$ differs from this value and by +0.14${^d}$ from the time of \cite{Sokolovsky2019}. This means that outburst detection occurred before maximum brightness was achieved. At the same time, this gives an estimate of the accuracy of the parameter t${_0}$ for this event no worse than $\pm$0.14${^m}$. The outburst amplitude of 2019 was estimated by combining modified light curves; this leads us to estimate the visual brightness in a quiet state at about 16.3${^m}$. The scatter of the points of the three light curves at the final phase of the light decline can be estimated by the following quantities $\Delta$m<0.5${^m}$ and $\Delta$log(t-t${_0}$)<0.1. The brightness errors are mainly related to the difference between instrumental systems and variations in the emission spectrum. 

The classification procedure consists in the complete coincidence of the light curves with the prototype, which means that a shift along the axis of the outburst amplitude may be required, which is permissible within 2-3${^m}$. The predecessor of the nova is the binary system, where matter from one star through an accretion disk, which is a significant source of radiation, flows onto the second star, which outbursts. Depending on the orientation of the accretion disk with respect to the observer’s line of sight, the total brightness of the system will change. According to \cite{Warner1987}, the amplitude of this change is estimated at about 2${^m}$. 

Our study focused mainly on visual and photographic observations: data for old novae are often mainly presented only by photographic observations. Visual observations of amateurs significantly complemented the ones of professionals. Since the end of the 20th century, CCD photometry of amateur astronomers has been actively developing. The multiplicity and the frequency of such observations made it possible to obtain a rather detailed light curve. A specific example of the sufficiency of dense series of visual observations of many observers is the V1974 Cyg light curve \citep{Kafka2019}: in the interval JD 2448700-2449200, the data are grouped along two parallel curves shifted along the amplitude axis. The reason for such an interesting manifestation of the features of the observational data lies in the systematic difference between the instrumental photometric systems of observers. At the final phases of the outburst, when the stars were significantly weaker and inaccessible to amateur astronomers, we used data in the photometric band V from publications, including the constantly updated database of the SMARTS project \citep{Walter2012}. In addition to the listed sources of observational data, modern outbursts of recurrent novae were tracked according to reports in urgent publications of the Circulars IAU, The Astronomer’s Telegram and similar analogues, which is indicated additionally. If necessary, on the basis of multicolor photometry data, estimates of the color index were performed and the resulting visual light curve was obtained after appropriate corrections if the data belong to other photometric bands (R, R${_C}$, I and I${_C}$). 

Historical outbursts have a lot of data from photographic surveys of the sky, which is equivalent to photometric band B. Comparison of the light curves in the photographic (or B) and visual (or V) bands did not show significant differences in the shape of the light curve (see, for example, Fig. 20 of \cite{Schaefer2010a} or numerous graphic material of \cite{Walter2012}). Modern surveys, for example, the OGLE project \cite{Mroz2016b}, make extensive use of observations in band I, in which the light curve can noticeably differ from the visual one; for example, V1141 Sco according to AAVSO visual data and OGLE I-photometry \citep{Mroz2015}. Extensive multicolor photometry of \cite{Walter2012} indicates that the V-I color index have different behavior at the maximum brightness and in later phases of the outburst; the amplitude of the changes can reach 1.5${^m}$. The final brightness decline on the light curve in the I band in the modified scales can begin at a different point in time compared to the V band, so this part of the light curve can be shifted relative to the visual light curve. The outburst amplitude may also be different, which will also require an additional shift to align the light curves. Therefore, some caution should be exercised here. 

When comparing light curves in different photometric bands, it must be understood that the brightness in a given band is the sum of the radiation in the spectral lines and in the continuum. If the photometric band is wide enough (several hundred angstroms), then two bands in the optical region of the spectrum will give light curves that differ little from each other. Since the 30s of the twentieth century, it is known that the emission lines observed in the spectrum of a nova, especially at the nebular, phase of a outburst increase the brightness of a star by several magnitudes (see, for example, \cite{Vorontsov-Velyaminov1953}). In this case, the light curves for the continuum region and for the region with strong emission lines will differ significantly \citep{Kaler1986}: the brightness in the continuum declines much faster than the brightness in the photometric band containing intense wide emission lines. For the 2010 U Sco outburst, there is multi-band photometry \citep{Pagnotta2015}, which indicates that there are no differences in the light curves in bands V and I. At the same time, T Pyx with a possible ejection of a small shell \citep{Duerbeck1979} display a slight difference in the light curves in the bands of V and I: in the middle part of the light decline there is no plateau in band I and the light decline branch is thus almost linear \citep{Walter2012}. Therefore, when considering such objects, it will be necessary to take this factor into account. A critical condition for the successful classification of a nova is the presence of a light curve throughout the outburst from the maximum brightness to a return to the quiet state. But there are situations with sufficient data when only a few observations (including the last pre-flare observation) are distributed over key parts of the light curve. 

Another problem arose when constructing the light curve using CCD photometry: when measuring images, we mainly used aperture photometry with annular diaphragms, within which, under a dense stellar field, faint background stars, including variables, can fall. This is especially true for photometric surveys of the Magellanic Clouds. The SMARTS review team \citep{Walter2012} warns about this and is working on a problem. In this situation, we preferred the data of the OGLE survey \citep{Mroz2016b}, since this specialized review was originally focused on a dense stellar field, the original measurement technique was used and it has already passed a long period of testing in other studies. 

For our further analysis, we used the light curves of the novae without any averaging, since each nova could have its own identities that would be noticeable when averaging, but they were hidden in the “noise” of a light curve, and this helped us from seeing the similarities we were looking for.

\section{Light curves of known recurrent novae}\label{sec3}

If the study by \cite{Schaefer2010a} does not see differences between the light curves of recurrent and classical novae, then in our modified version of the presentation of light curves recurent novae differ markedly from classical novae. From the standpoint of this study, in the presence of a complete light curve, a nova outburst can be reliably attributed to one or another type. The first sign that allows us to suspect a recurrent type of outburst of a nova is traditionally attributed to the amplitude of the outburst \citep{Duerbeck1987b}: in recurrent novae, it does not exceed 11m at higher values in classical ones. The outburst duration of a recurrent nova is also noticeably shorter than that of a classical nova. Therefore, the light curve of the recurrent nova in modified scales has a noticeably faster brightness decline after principal maximum or the greater slope of the final brightness decline (see also the second part of this review). Below we will turn to the search for candidates for recurrent novae among the classical novae. But first, consider 10 well-known recurrent novae (Table 1), for which an extensive study of \cite{Schaefer2010a} presents data on re-measurements of photographic archives and smoothed modern dense series of visual and other observations, which we will supplement when necessary from other sources. 

Table 1 contains the main outburst parameters of recurrent novae necessary to reproduce their light curves in modified scales. They are enough for the practical reproduction of the light curves of known novae in a series of outbursts and for their subsequent comparison with a light curve of unknown or erupted recently nova in order to determine the type of its outburst. Hereinafter, the parameter t${_0}$ will be determined as the difference of the Julian date (JD) of the moment of the outburst maximum as it was determined by us in Section 2 and the date JD=2400000: t${_0}$=JD-2400000(=nnnnn.n) or t${_0}$=JD-2450000(=nnnn.n) for the present day. The amplitude A of the outburst as well as the current brightness of a nova will be determined relative to the accepted magnitude of the pre- or post-nova m${_q}$; usually it will be in the visual, or in the photometric band V, the brightness scale, or we will additionally specify when these will be other photometric bands: p – photographic, B, R, I, and others. The fifth column of Table 1 contents the average B, V, and R magnitudes of the pre-/post-nova from Table 25 of \cite{Schaefer2010a}. The absent parameters of Table 1 for some outbursts were not determined due to an insufficiently complete light curve. Others abbreviations are: V$_{max}$ - the brightest or peak V-band magnitude or in other photometric bands, V$_{q}$ - the quiescent V- or other band magnitude, M$_{V}$ - the absolute maximal magnitude in the V- or other bands.

\subsection{CI Aql}

Three outbursts of this star are known: 1917, 1941 and 2000. In the review of \cite{Schaefer2010a}, photometric data on outbursts are substantially supplemented. A detailed visual light curve is available for the last outburst, the first two, photographic light curves are less detailed, but there are no noticeable differences (Fig.1). The parameters of expression (2) for constructing light curves in Fig.1 are given in Table 1. 

The CI Aql light curve at the maximum light state has a small plateau with the duration of 8–9 days. A characteristic detail is the subsequent wave-like decline in brightness. 

\begin{center}
	\begin{table}[t]%
		\centering
		\caption{Parameters of the light curves of recurrent novae in the outburst amplitude scale and the logarithmical scale of time. \label{tab1}}%
		\tabcolsep=0pt%
		\begin{tabular*}{250pt}{@{\extracolsep\fill}rccr@{\extracolsep\fill}}
			\toprule
			\textbf{Nova, year$ $ $ $ $ $ $ $ $ $ $ $}& \textbf{t${_0}$,}& \textbf{A/m${_q}$}& \textbf{B, V, R $ $ $ $ $ $ $ $ $ $ $ $} \\
			\midrule
			CI Aql    1917& $\geq$21365 & 8/16.7p & 17.17,16.16,15.51 \\
			1940 & 30060$\pm$10 & >3.4/17.2p & \\
			2000& 51661  & 7/16.2 & \\ 
			V394 CrA  1949 &32998 &>10.3/17.8p  & 18.6,18.0,17.5  \\ 
			1987 & 47005.5 & 11.5/18.5 & \\ 
			T CrB     1866 &02734.2  &7/9.5 & 11.57,10.04,8.95\\ 
			1946 &31860.3 &7/9.75  & \\ 
			IM Nor    1920 &22495 &9.4/19.35p  &19.35,18.36,17.77  \\ 
			2002&52270 &10.7/18.5  & \\ 
			RS Oph 1898,1907,&- &- & 12.23,11.05,10.25 \\ 
			1933,1945,1958,&- &- & \\ 
			1967&39790.5  &7.1/12.0  & \\
			1985&46093.9  &7.1/12.0 & \\
			2006&53779.6 &7.1/12.0 & \\
			V2487 Oph 1900 & - & - &18.11,17.34,16.67 \\ 
			1998&50979 &7.3/17.5  & \\ 
			T Pyx     1890 & - & - &15.51,15.40,15.24  \\
			1902&15842 &9.0/16.0p  & \\
			1920&22390$\pm$3 &9.5/17.0p  & \\
			1944&31390$\pm$10  & 9.5/15.8p  & \\ 
			1967&39468$\pm$1 &9.6/15.7  & \\ 
			2011&55668$\pm$0  &9.5/15.8 & \\
			V3890 Sgr 1962 &37808$\pm$1  &>5.7/16p  &16.26,15.33,14.12  \\
			1990& 48008$\pm$1  & 8/16  & \\
			2019& 58723.21  &9.6/16.3 & \\ 
			U Sco     1863 & 01644 &9.3/18.4  & 18.5,17.6,17.3  \\ 
			1906, 1917& - & - & \\
			1936& 28341.45  &10.6/19.0p & \\ 
			1945& 31604 &10.0/18.8p  & \\
			1969& 40250  &>5.7/18.5  & \\ 
			1979& 44046.4 & 9.8/18.5 & \\
			1987& 46929 & >8.2/19.5 & \\ 
			1999& 51234.82 & 10.8/18.5 & \\
			2010& 55224.6  & 10.6/18.5 & \\ 
			V745 Sco  1937 & 28663.5 & 10.3p/20.2p  & 20.5,18.8,16.2  \\
			1989& 47736.4  & >8.7/18.3 & \\ 
			2014& 56695.1 & 10.3/19.0  & \\
			\bottomrule
		\end{tabular*}
	\end{table}
\end{center}

\begin{figure}[t]
	\centerline{\includegraphics[width=78mm]{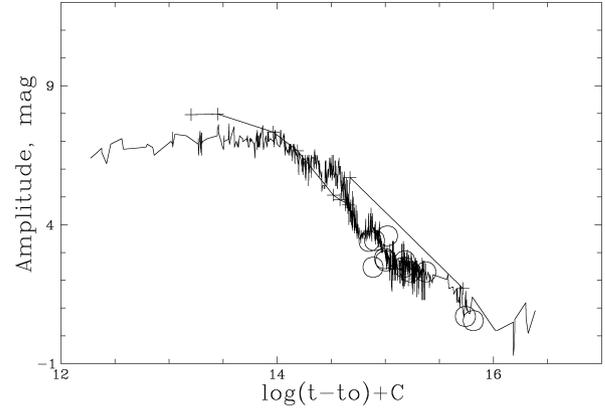}}
	\caption{Light curves of CI Aql in a logarithmic scale of time. (Hereinafter, the ordinate scale starts from -1 to display the star’s brightness reaching the level of quiet state.) The broken line with pluses is the outburst of 1917, the circles are the data for 1940, the broken line is the outburst of 2000.\label{fig1}}
\end{figure}

\subsection{V394 CrA }

Two outbursts of 1949 and 1987 are known (Table 1). The light curve is very incomplete (Fig.2). Apparently, the maximum brightness is short: the increase in brightness is replaced by a decline over a time of the order of a day. Next, a uniform decrease in brightness occurs, which accelerates near the level of 7.5$^{m}$ and slows down again at the level of 5$^{m}$. The shape of the light curve at the maximum was reconstructed by choosing the moments t${_0}$ for the linearization of the light curve of the same type at the phase of the initial decline (see details in Section 3.10 for V745 Sco). 

\begin{figure}[t]
	\centerline{\includegraphics[width=78mm]{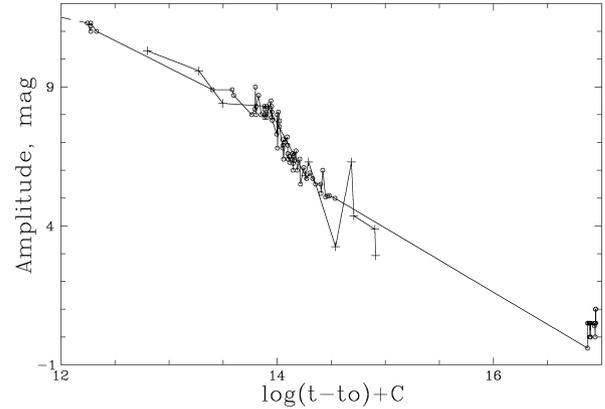}}
	\caption{V394 CrA light curves in a logarithmic scale of time. The broken line with pluses is the outburst of 1949, the broken line with circles is the one of 1987. \label{fig2}}
\end{figure}

\subsection{T CrB}

Figure 3 shows the light curves for the known outbursts of 1866 and 1946 with the parameters of Table 1. The initial rise was fast, no more than two days. No more than this was the duration of the state of maximum. The decline in brightness tended to accelerate on a logarithmic time scale, and near a brightness of 2.4$^{m}$ above a quiet level, the speed of decline in brightness began to slow down; as noted above, this may be due to an increase in the contribution of the second radiation source to the overall radiation of the system. The light curve was smooth: without plateau/stop details. A remarkable detail of the two T CrB outbursts is a repeated increase in brightness with an amplitude of up to 2$^{m}$ at log(t-t${_0}$)+C$\approx$15.15. The energy emitted in the visual region of the spectrum in this secondary outburst is only an order of magnitude less than the energy emitted in the main outburst. 

\begin{figure}[t]
	\centerline{\includegraphics[width=78mm]{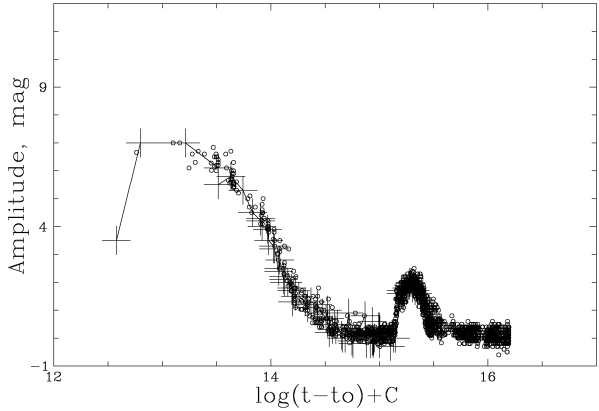}}
	\caption{Visual light curves of T CrB in a logarithmic scale of time. Line with pluses - outburst of 1866, circles - outburst of 1946. \label{fig3}}
\end{figure}

\subsection{IM Nor}

Light curves of two outbursts of 1920 and 2002 are presented in Fig.4 with the parameters of Table 1. As follows from the data of \cite{Schaefer2010a} for the 1920 outburst, the maximum duration was at least 29 days. In 2002, the state of the maximum lasted up to 20 days (\cite{Brown2002, Schaefer2010a}). Therefore, a prolonged state of maximum brightness is a characteristic feature of this nova. The brightness decline can be represented by two straight segments with a short plateau near the brightness level of 4-6$^{m}$. In the first section, the observations are quite numerous, which allows us to confidently say that there are no details in the behaviour of the star. The same can be allowed for the second section with its rare observations, since the representation of it by a straight line does not cause great objections. In Fig.4, CCD observations without a filter \citep{Schaefer2010a}, which can be considered as observations in the photometric R band, are corrected for the color index V-R=0.59$^{m}$. 

Here, one can pay attention to the difference between episodic estimates of brightness of the pre-nova by \cite{Kato2002}: B=18.0$\pm$0.3$^{m}$ in 1975, V=19.4$\pm$0.4$^{m}$ in 1987, R=17.0$\pm$0.3$^{m}$ in 1992, data of \cite{Schaefer2010a} (see Table 1): B=19.35$^{m}$, V=18.36$^{m}$, R=17.77$^{m}$, and photometry by \cite{Woudt2003}: V=16.5$^{m}$. Taking into account Fig.4, it becomes clear that the last photometry refers still to the active state, and until it is completely returned to a quiet state, remains still almost 6–9 years and the brightness will decrease still by almost 2$^{m}$. Consequently the eclipse depth in the IM Nor system can be significantly higher than the recorded 0.2$^{m}$ by \cite{Woudt2003}. Such the long outburst duration is more typical of the classical nova, along with a possible ejection of matter during the outburst.

\begin{figure}[t]
	\centerline{\includegraphics[width=78mm]{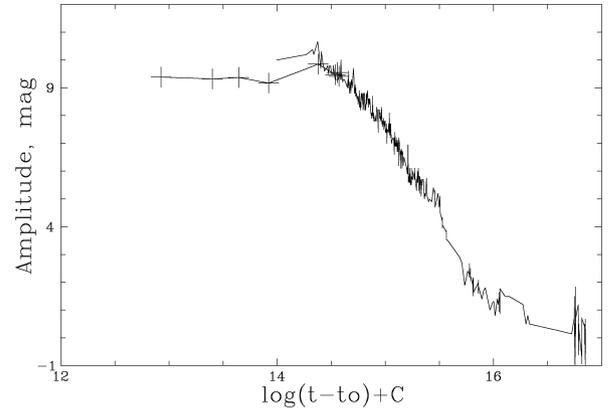}}
	\caption{Light curves of IM Nor in a logarithmic scale of time. The broken line with pluses is the outburst of 1920, the broken line is the outburst of 2002.\label{fig4}}
\end{figure}

\subsection{RS Oph}

A small recurrence time (over 118 years of 8 outbursts) made it possible to “catch” three RS Oph outbursts in a timely manner and obtain a detailed light curve (Fig.5, Table 1). The summary light curve was obtained without any difficulties: the instant of maximum brightness t${_0}$ was unambiguously recorded, the outburst amplitude was always the same and was counted from the brightness level to which the star returned immediately after the outburst (and then there occurred out a slow brightness increase of about 1$^{m}$ and the star in such remained until the next outburst). The initial rise of brightness during a outburst takes from 1 day (1933) to 2 days (1958 and 1985). This is followed by a monotonous decrease in brightness to a level of 2-3$^{m}$ above the quiet state, after which the decrease in brightness is accelerated and the star weakens to a level of about 1m below the average brightness level in a quiet state. The last details are clearly visible and described in the traditional representation of the light curve (see Fig. 24-31 of \cite{Schaefer2010a}), a brightness level of about 2$^{m}$ corresponds to a plateau according to \cite{Schaefer2010a}.

\begin{figure}[t]
	\centerline{\includegraphics[width=78mm]{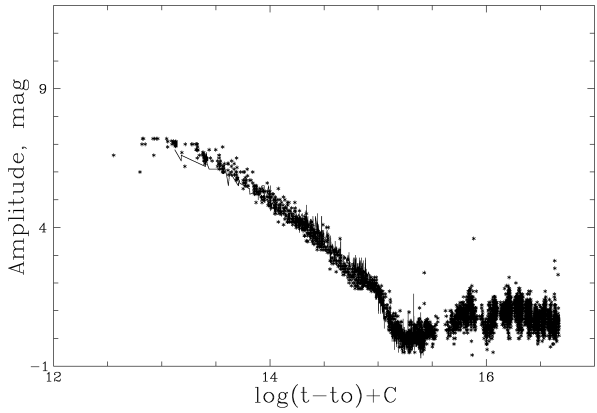}}
	\caption{Light curves of RS Oph in a logarithmic scale of time. The broken line is the outburst of 1985, the dots are 1967 and 2006.\label{fig5}}
\end{figure}

\subsection{V2487 Oph}

A star that outburst in 1998 was classified as a recurrent one in 2009 after archival detection of the 1900 outburst (single plate with double exposure) \citep{Pagnotta2009}.

\begin{figure}[t]
	\centerline{\includegraphics[width=78mm]{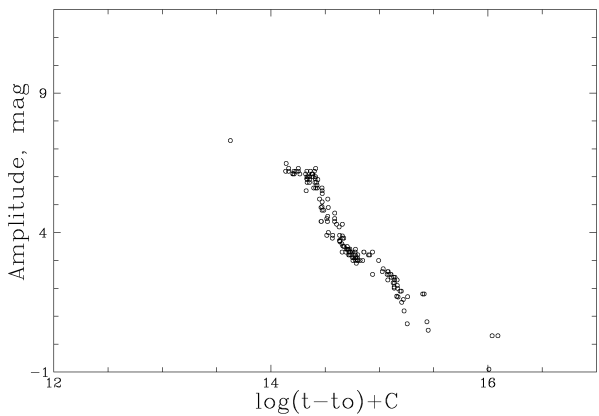}}
	\caption{Light curve of V2487 Oph in 1998 on a logarithmic scale of time.\label{fig6}}
\end{figure}

\subsection{T Pyx}

The parameters of light curves (Fig.7) in some of the known T Pyx outbursts are given in Table 1. In the state of maximal brightness, the visual ($\equiv$V) and photographic ($\equiv$B) light curves do not differ. The outbursts of 1967 and 2011 are well represented by visual observations and allow us to note some differences. In particular, on the branch of brightness decline near the 4$^{m}$ level there is an inclined plateau with a brightness decline of about 1$^{m}$. The reality of such a difference is beyond doubt, since the value of the brightness difference is higher than the scatter of data and prior to this plateau, the light curves had the opposite position: the brightness in 1967 was slightly higher than in 2011, and after the step they exchanged places. There is also a difference between the outbursts of 1902 and 1967 from 2011 at the beginning of the state of maximum. The 1967 outburst is also notable for the increased brightness on the last dates of observations. But, apparently, the reason for such an increased level of brightness is associated with observation errors near the limit level for visual observations. 

The outburst of 2011 has a very well-presented light curve in the AAVSO database. This allowed us to get an idea of the behaviour of the nova brightness from the start up to the end of the outburst. The time moment t$_{0}$ determined according to Section 2 also retains its significance for obtaining the light curve in the ascending part of the outburst to the maximum. Expression (2) is transformed to the form 

log(r)=-log(t$_{0}$-t)+C.   $ $ $ $ $ $ $ $ $ $ $ $ $ $ $ $ $ $ $ $ $ $ $ $ $ $ $ $  (3)

On the resulting light curve, the points near the t$_{0}$ moment mutually overlap and, for the convenience of consideration, they were discarded. It is seen that the brightness of the quiet state before the outburst is slightly higher than after the outburst. The increase in brightness in the logarithmic time scale occurred almost linearly. The state of the maximum of outburst begins as a plateau with a slight trend for an increase in brightness for about 14 days, which ends with a rapid increase in brightness by about 1$^{m}$. This rise is the beginning of 2-3 ''pulsations'' of brightness with the amplitude of about 1.2$^{m}$ (within about 20 days) and the nova goes into the phase of brightness decline.

\begin{figure}[t]
	\centerline{\includegraphics[width=78mm]{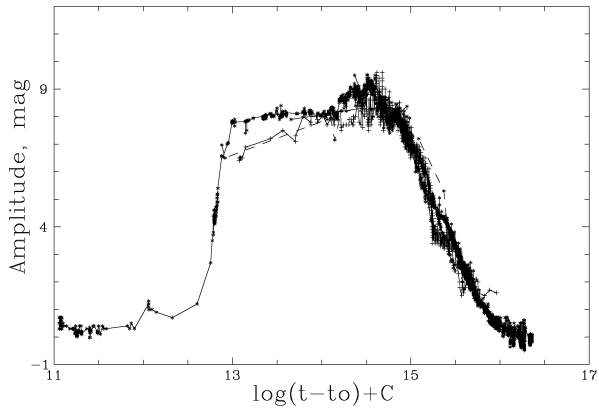}}
	\caption{Light curves of T Pyx in a logarithmic scale of time. The dashed line with dots is the outburst of 1902, the line with pluses is the data for 1967, the dotted line is the outburst of 2011. The light curve for 1920 is within the three above specified light curves. For an explanation of the pre-maximum portion of the light curve see text. \label{fig7}}
\end{figure}

\subsection{V3890 Sgr}

Light curves for three well-known outbursts of 1962, 1990 and 2019 are shown in Fig.8. As we noted above, the outburst of 2019 was a kind of test of our methodology. 

The parameters of Table 1 were obtained before the 2019 outburst by combining the light curves of the first two outbursts due to a distinctive feature: the plateau near a 5$^{m}$ level. The t$_{0}$ parameter for the 1962 outburst was chosen to better match of these two the light curves (for details see below the section for V745 Sco). It also follows from these parameters that in 1962 the maximum brightness was missed, and the second light curve was almost whole: the outburst, apparently, was detected no later than a day after reaching the maximum brightness. 

Observations in 2019 provided an obvious knowledge of the parameter t$_{0}$. This value provided an excellent combination of all three modified light curves, and also meant the correctness of our methodology for constructing modified light curves.

\begin{figure}[t]
	\centerline{\includegraphics[width=78mm]{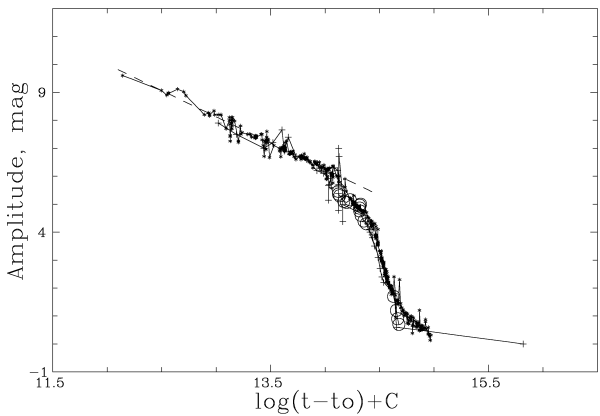}}
	\caption{Light curves of V3890 Sgr in a logarithmic scale of time. The line with circles is the outburst of 1962, the line with pluses - 1990, the line with dots – 2019. The dashed straight line is the linear approximation of the initial decline phase with the slope of -1.875.\label{fig8}}
\end{figure}

The dense series of high-precision observations from two data-base, AAVSO and ASAS-SN, for the 2019 outburst repeated each other in detail, so we can talk about the features of the process of recovery a binary system to its normal state. The transition to a faster decline in brightness near log(t-t$_{0}$)+C$\approx$14.05 coincides with the start of the super-soft source (SSS) emission \citep{Page2019}. The plateau on 5$^{m}$ for 1961 and 1990 is absent for 2019. Apparently, at this time, there was a significant variability in the intensity of the line spectrum of the nova, which was manifested in a sharply increased scatter of brightness estimates by different observers with different instrumental systems: visual or with a broadband filters. Differences in the light curves of the outbursts of 1962 and 1990 from the 2019 outburst after the bend point of log(t-t$_{0}$)+C$\approx$14.1 indicate a real difference in the conditions in the active binary system. The outbursts of 1962 and 1990 sharply “turned off” near log(t-t$_{0}$)+C$\approx$14.7, while in 2019 the decline in brightness, starting from a brightness level of about 2$^{m}$, gradually slowed down, and only near abscissa of 14.9, the brightness declined sharply to a quiet level of 16.3$^{m}$.

\subsection{U Sco}

After comprehensive research on outbursts of recurrent novae of \cite{Schaefer2010a}, the number of known outbursts of U Sco increased from 9 to 10: in 2010, another outburst occurred. In Fig.9, the light curves of 8 outbursts are combined with the parameters of Table 1, except for flares of 1906 and 1917 represented only by single measurements of brightness. 

Most outbursts showed similar light curves. The brightness rise did not exceed 24 hours, the state of maximum brightness lasted less than a day, and then there was a rapid decline in brightness. The brightness decline can be represented by 4 linear sections or waves around linear trend: from the maximum to 8$^{m}$, 8-4.5$^{m}$, 4.5-4$^{m}$ and 4-1$^{m}$. 

Data for the 1945 and 1969 outbursts do not go beyond the “noise” track in Fig.9, so they are not shown in the figure. The initial linear decline of the summarized modified light curve has a negative slope of $\geq$-2.2. For a confident reconstruction of the light curve in the maximum region, the accuracy of time fixing for the light estimates should be at the level of hundredths of a day and better and, accordingly, a higher frequency of the follow-up of the light estimates is necessary. Currently, only three outbursts (1936, 1999, and 2010) have brightness estimates near the maximum.

\begin{figure}[t]
	\centerline{\includegraphics[width=78mm]{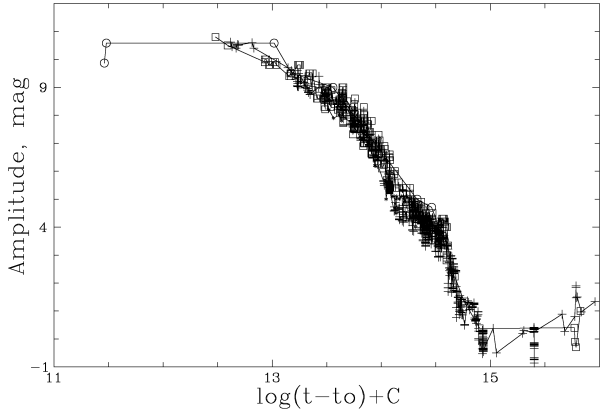}} 
	\caption{Light curves of U Sco in a logarithmic scale of time. The line with circles is the outburst of 1936, the line with dots - 1987, the line with squares - 1999, the line with pluses - 2010. \label{fig9}}
\end{figure}

\subsection{V745 Sco}

Three well-known V745 Sco outbursts formed a good idea of the shape of the light curve despite the different photometric bands used in its construction: in 1937 - photographic data, or band B, (see compilation of \cite{Schaefer2010a}), 1989 - out-system \citep{Liller1989}, visual \citep{Kafka2019} and photometry in the V band \citep{Sekiguchi1990}, 2014 - visual \citep{Kafka2019}, photometry in the bands of V \citep{Walter2012} and I \citep{Mroz2014, Mroz2015}. The parameters for light curves are presented in Table 1. The combining of the curves in terms of the amplitude of the outburst was carried out on the transition from the first linear part of the light curve to the second. There are no observations when the nova brightness reaches the level near quiet state due to the great weakness of the object. 

The increase in brightness to a maximum was fast and took place over a period of about 1 day. The maximum brightness may have a very sharp peak: the initial brightness decline of 1.4$^{m}$ occurred in no more than 1 day, while the subsequent decline of 1.4$^{m}$ took already 4 days. In the 1989 outburst, this peak was probably missed by observations. The linear sections of the initial brightness decline of all three outbursts, i.e. the t$_{0}$ parameter were restored with an orientation on the 2014 outburst. As similar situations with some other fast recurrent novae, V394 CrA and V3890 Sgr, showed, the slope of the linear initial brightness decline has a coefficient of about -2 or L$\propto$r$^{-0.8}$.

\begin{figure}[t]
	\centerline{\includegraphics[width=78mm]{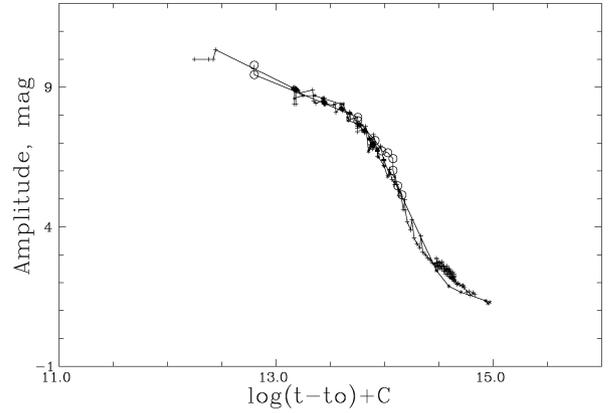}}
	\caption{Light curves of V745 Sco in a logarithmic scale of time. The line with circles is the outburst of 1937, the line with dots - 1989, the line with pluses - 2014. \label{fig10}}
\end{figure}

\section{Discussion}\label{sec4}

The summary light curves of some of the novae give a very good idea of all the phases of the outburst. This is, first of all, a fast recurrent nova of U Sco and a slower recurrent nova of RS Oph with their numerous outbursts and high maximum brightness, which made these stars accessible to many observers. T CrB, CI Aql, V745 Sco, and IM Nor are represented by dense series of observations of one outburst. At the same time, the light curves of V394 CrA, V2487 Oph and V3890 Sgr only now outline a characteristic general appearance. But the most complete picture from the start to the end of the outburst take place for the T Pyx light curve, which, thanks to the brightness in a quiescence state, is available for constant patrolling with small telescopes, while the slow development of the outburst allowed us to have a good end result. 

From a methodological point of view, the stability of the specific shape of the light curve to the accuracy of determining the moment of maximum is clearly visible. The latter circumstance comes to the fore in the case of fast novae (V745 Sco). Here it is already very desirable to have denser series of observations or set of data for several successive outbursts and then adhere to uniformity in the iteration of the maximum moment, which on the light curve in the modified axis of abscissa should not deviate up and to the right from the subsequent linear shape of the light curve. 

For 8 recurrent novae, the light curves constructed according to our procedure have a maximum or the beginning of a maximum brightness state at log(t-t$_{0}$)+C=13.1. But for fast outbursts of V394 CrA, V745 Sco, and etc, the maximum brightness is shifted to the left of this point in order to eliminate the up of the light curve in the modified time scale above the linear approximation of the subsequent trend, as we mentioned above in the description of our methodology. A similar situation will arise in the future when the maximum brightness is very sharp and the time t$_{0}$ will be selected with an accuracy of a fraction of a day [the light curve of the classic new V1500 Cyg with a dome-shaped maximum is indicative in this respect]. In the future, this formalism, when the light curve is aligned in the maximum region, will help us in grouping classical novae. 

Obvious characteristic details are also visible in the light curves of recurrent novae: a plateau that precedes the brightness maximum; the presence of linear sections; stops or plateau in a decline of brightness, or wave-like changes in a decline of brightness at the final phase of a outburst development. The final phase of the light curve can be an almost straight line to the level of quiescence brightness (U Sco) or can end with a smooth monotonous curve (T Pyx, V745 Sco), starting at a level of about 2$^{m}$. 

Single detail can also be found in other novae, but this should be considered as an individual feature, if there are differences in other details. So, the slope of the final brightness decline is almost the same for U Sco and T CrB, but the amplitudes and state of maximum brightness differ. 

The main conclusion we draw from the review of modified light curves of novae in recurrent outbursts is the stable shape of the light curve, i.e. the repeatability of the main details of the outburst process: the shape of the light curve at maximum or in the state of maximum brightness, the slope of the light curve in the initial decline phase and the presence of wave-like variations. For some stars, this conclusion was based on sequences of 4-5 episodes with numerous observations. The light curve of the V3890 Sgr in the third 2019 outburst showed this very well. A dense series of observations made it possible to determine the parameter t$_{0}$ already in the first 3-5 days (Table 1), and subsequently we only monitored the repetition of all deviations from linearity that were known from the outbursts in 1962 and 1990. In other words, this allows us to see critical details of the light curve behaviour, and in the future to more closely trace these outburst points. The well-known result that the maximum brightness in the outbursts is stable was confirmed, i.e. during the outbursts of a nova, the same energy was released each time. It is also important that basically is confirmed the (3) conclusion of \cite{Schaefer2010a}: “All the eruption light curves from a single RN are consistent with a single invariant template. That is, the light curves are always the same from eruption to eruption. This tells us that the eruption light curve depends on system parameters (like the white dwarf mass and the composition) that do not vary from eruption to eruption (like mass loss rate)". The above noted (Section 3.7) small differences between the T Pyx outbursts in 1967 and 2011, it can be more likely attributed to the existence of a real range of admissible variations of the binary system parameters, which are critical for triggering a thermonuclear process.  

These conclusions create the basis for the search for analogues among other novae, primarily among the known members of the existing small group of recurrent novae. We try to imagine the order among the light curves of recurrent novae which we presented in our preliminary study \citep{Rosenbush2002} and which \cite{Schaefer2010a} described as “an extreme example of the chaos and instability in RN subdivisions”. We will turn to classical novae with a similar search after acquiring some experience in analysing this group of novae.

\subsection{Groups of recurrent novae}

To reduce doubts due to the possible difference in photometric scales due to differences in comparison stars and other reasons \citep{Schaefer2010a}, we compared only the last outbursts of each star in the modern era. Here, amateur astronomers with very dense series of observations at all phases of the outbursts, from the start of an outburst to its finish, made an invaluable contribution. 

Common signs may include: the presence of a plateau at the maximum, the slope of the light curve in the initial decline phase, the variability of light curve during the transition phase of the outburst or the presence of other peculiar details of the light curves. The main criterion for grouping is the coincidence of light curves (a visual assessment of the degree of coincidence at this phase of present study). 

In this section, we will form groups among the known recurrent novae, and then we will form the list of the possible recurrent novae among the known novae with single outbursts, for the most part these are classical novae. Compared to the preliminary study by \cite{Rosenbush2002}, two groups retained their composition: IM Nor (+ T Pyx) and CI Aql (+ V2487 Oph). We will begin our consideration with them. 

\subsubsection{T Pyx group} 

This group with the IM Nor prototype with its 2002 outburst was early highlighted by \cite{Rosenbush2002}, but here we will choose T Pyx as the prototype and justify this choice by the level of reliability of the summary and most complete light curve with 6 known outbursts versus 2 ones for IM Nor. The similarity of the light curves of IM Nor (Fig.4) and T Pyx (Fig.7) during the light curve decline indicates the possibility of combining them: the similarity of light curves includes the equality of the slopes of the decline after the maximum (Fig.11). One more argument in favour of combining is an extended plateau in the state of outburst maximum, which was always observed in T Pyx, and only once in IM Nor during the outburst in 1920 (in the second outburst, this phase is not covered by observations). The plateau, or state of maximum brightness, of T Pyx begins with a brightness level of about 8$^{m}$; after about 11 days (t$_{0}$ + 11$^{d}$), the brightness rises from the level of 8.2$^{m}$ to 9.2$^{m}$ and on the 25th day the plateau ends with a fast decline in brightness. In IM Nor, such changes took 20-30 days. 

\begin{figure}[t]
	\centerline{\includegraphics[width=78mm]{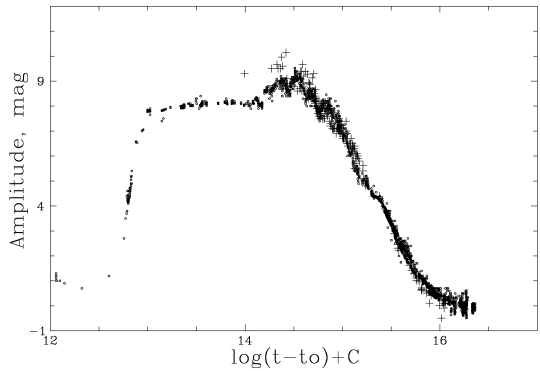}}
	\caption{Summary light curve for stars - members of the T Pyx group in a logarithmic scale of time. Line with pluses - data of IM Nor, points - T Pyx.\label{fig11}}
\end{figure}

To combine the light curves in Fig.11, the amplitude of IM Nor was reduced by 0.7$^{m}$. Such a procedure is admissible, since the brightness of the quiescence of a binary system with an accretion disk depends on the spatial orientation of the system with respect to the observer \citep{Warner1987}.  

\subsubsection{CI Aql group} 

The similarity of the light curves of V2487 Oph and CI Aql (Fig.12) served us earlier \citep{Rosenbush2002} as the basis for the assumption that the first star belongs to recurrent novae. Somewhat earlier, the same assumption was independently made in the study of \cite{Hachisu2002}, based on theoretical study. The outburst of this object in 1900 was detected on a single double-exposure photographic plate by \cite{Pagnotta2009} as a result of viewing archives of two photographic sky surveys. 

\begin{figure}[t]
	\centerline{\includegraphics[width=78mm]{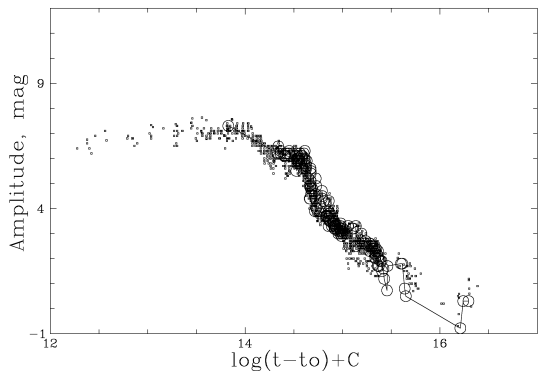}}
	\caption{Summary light curve for stars - members of the CI Aql group in the logarithmic scale of time. Dots - data of CI Aql, the line with circles - V2487 Oph.\label{fig12}}
\end{figure}

\subsubsection{V745 Sco group} 

In addition to the above two groups of recurrent novae, the third group was allocated with a base on the V745 Sco (Fig.13). 
In this group, we included V394 CrA and V3890 Sgr. To combine the light curves of V745 Sco and V394 CrA in Fig.13, the amplitude of the outbursts of the latter was changed on -0.6$^{m}$, the amplitude of V3890 Sgr was changed on +1$^{m}$. 

These three stars combine two details: the similarity of the initial brightness decline and the transition to the final decline phase near log(t-t$_{0}$)+C$\approx$14.0. Close to the finish of the final decline phase, the light curves occupy a strip of width $\Delta$log(t-t$_{0}$)$\le$0.3 and $\Delta$A$\le$2.5$^{m}$. Outbursts have relatively close durations. 

Thus, outside these three groups, 3 recurrent novae of 10 known remained: T CrB, RS Oph, and U Sco. The light curves are very detailed, so there is no doubt about the uniqueness of the stars. But each of these stars has close analogues among possible candidates for recurrent novae, that will be discussed below. U Sco and V3890 Sgr have similar details, but to combine them, a shift is required, both in amplitude and on the abscissa. The shift on the abscissa is interpreted by us as belonging to another group of novae. The shift along the amplitude within 2$^{m}$ may be due to the spatial orientation of the binary system \citep{Warner1987}; the negative shift may mean less luminosity of the original binary system. 

\begin{figure}[t]
	\centerline{\includegraphics[width=78mm]{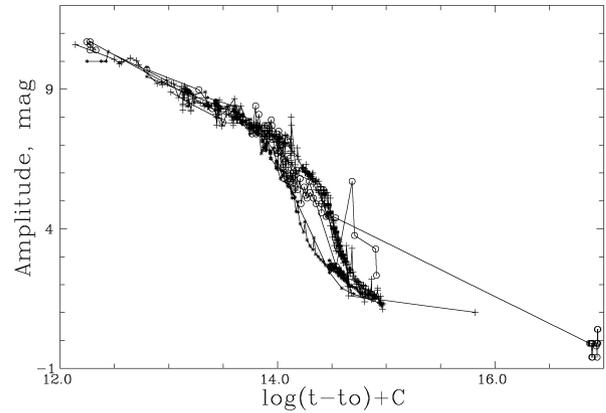}}
	\caption{Summary light curves of novae - members of the V745 Sco group. V745 Sco - three broken lines with dots, V394 CrA - two lines with circles, V3890 Sgr - three lines with pluses.\label{fig13}}
\end{figure}

\subsection{Recurrent novae and candidates for recurrent novae in the Magellanic Clouds}

The modern instrumental capabilities of observations of novae of nearby galaxies have reached an impressive level, which allows us to quickly replenish the observational database. The permeability of the robotic complexes (the detection limit reaches I>21.5$^{m}$) and their operability allows almost completely covering a nova outburst from the beginning of the process to the complete completion. This is especially true for photometric surveys of the Large (LMC) and Small (SMC) Magellanic Clouds. Recent reviews of the properties of novae in the LMC and SMC are presented in studies of \cite{Pietsch2010, Shafter2013, Mroz2016b}. Moreover, the latest publication cites detailed photometry data that covers almost the full amplitude of the flares and allows us to fairly accurately estimate the moment t$_{0}$. This allows us to confidently classify the light curves of novae outbursts, including belonging to one groups of the recurrent novae identified above or creates  one more group with a nova of three unique ones (T CrB, RS Oph, or U Sco). 

It is known 3 recurrent novae in the LMC: LMCN 2004-10a, LMCN 2010-11a \citep{Mroz2014} and LMCN 2009-02a \citep{Bode2016}. Our review of 13 novae of the LMC from Atlas of \cite{Mroz2016b} revealed that at least 3 of them are recurrent in addition to the already known three (Table 2). Among 9 novae of the SMC, 4 objects were assigned to candidates for novae (Table 2). Table 2 shows the closest match to the groups or specific recurrent novae of our Galaxy. For the brightness of a quiet state, only values derived from a comparison with the summary light curve with prototypes are given; it is problematic to say how much these values differ from the actual brightness, since they are at the level of the limit of 21.5$^{m}$ for the OGLE survey \citep{Mroz2014, Mroz2016b}. We did not perform offsets on the abscissa scale in order to have an idea of errors in the classification; recall that with distance from the maximum moment the shape of the light curve is practically unchanged. 

Fig.14 shows the light curves of recurrent novae and candidates for recurrent novae of the T Pyx group, the most easily identified group. 

The SMCN 2001-10a was discovered by \cite{Liller2001} and had several non-system estimates of brightness of \cite{Liller2004b}, which were insufficient to determine the shape of the light curve. An almost complete light curve of \cite{Mroz2016b} leaves no doubt to include this nova in the candidates for the T Pyx group with the parameters of Table 2 and the duration of the state of maximum brightness for about 15 days. It note that this is only 4 days longer than the estimate of time of maximum based on the state of the spectrum in the study of \cite{Mason2005}. It is interesting to note that within 1-2 days after the state of maximum brightness, the color index v-r increased from about 0-0.3$^{m}$ to 1.0$^{m}$ \cite{Liller2004b}. 

SMCN 2006-08a was observed only in the OGLE survey \citep{Mroz2016b}. The light curve (Fig.14) is presented almost completely, with the exception of the maximum, which could take place no more than 8 days after the last observation of nova in the quiescence, i.e. in the interval JD 2453950.82-2453957.78. The similarity of T Pyx is not in doubt. In Fig.14, the first point of the light curve displays the possible part of the light curve until the first positive observation of this nova.

\begin{figure}[t]
	\centerline{\includegraphics[width=78mm]{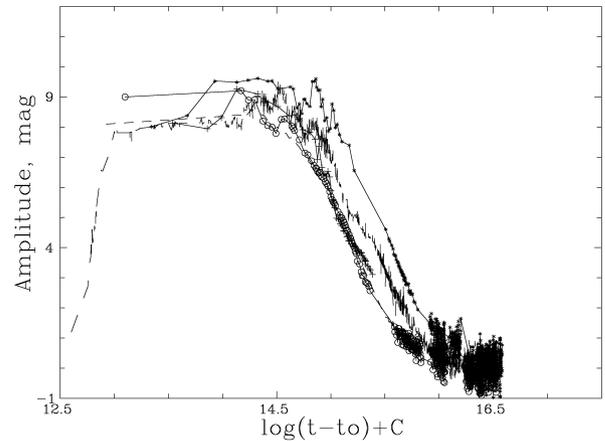}}
	\caption{Candidates for the recurrent novae of the Magellanic Clouds (Table 2) - members of the T Pyx group: dashed line is the light curve of T Pyx in 2011, line with dots - LMCN 2005-11a, line with pluses - LMCN 2013-10a, dashed line - SMCN 2001-10a, line with circles - SMCN 2006-08a. \label{fig14}}
\end{figure}

\begin{center}
	\begin{table*}[t]%
		\caption{Some parameters of the known recurrent novae and candidates in the recurrent novae of the LMC and SMC according to this study.\label{tab2}}
		\centering
		\begin{tabular*}{400pt}{@{\extracolsep\fill}rlclcc@{\extracolsep\fill}}
			\toprule
			\textbf{Nova} & \textbf{t$_{0}$}  & \textbf{I$_{max}$/V$_{max}$}  & \textbf{Novae group} & \textbf{M$_{Imax}$/M$_{Vmax}$}  & \textbf{M$_{V}$} \\
			\textbf{}     & \textbf{}          & \textbf{}                       & \textbf{or nova/I$_{q}$}    & \textbf{}           & \textbf{\citep{Shafter2013}} \\
			\midrule
			LMCN 2004-10a & 3298.2     & 11.65/10.8  & RS Oph/19.5   & -6.7/-7.9   & (-8.3)-(-8.0)    \\
			1937-11a      &            &             &               &             &                  \\
			LMCN 2005-11a      & 3692  & 11.4/11.5  & T Pyx?/21   & -7.2/-7.2  & -7.4    \\
			LMCN 2009-02a     & 4867  & -/11.4:  & RS Oph/19.75 & -/-7.3  & -8.3    \\
			1971-03a     &   &   &   &   &     \\
			LMCN 2016-01a     & 7408.5  & 11.5/12.2  & CI Aql/19.8   & -5.9/(-7.5)-(-8.5:) 	  & -8.7    \\
			2010-11a     & 5522  & 12.7/-  &   &   &     \\
			2002-02a     &   &   &   &   &     \\
			1990-02a     &   & -/10.2  &   &   &     \\
			1968-12a     &   & -/10.9p  &   &   &     \\
			LMCN 2012-03a     & 6012.4 & -/11.9  & T CrB/18.17  & -/-6.8  & <-8.2:    \\
			LMCN 2012-10a     & 6223  & (9.5)I/-  & V745 Sco/20I  &   &     \\
			LMCN 2013-10a     & 6579  & 12.4/-   & T Pyx/20.5   & -7.4/-   &     \\
			SMCN 2001-10a     & 2192  & 11.9/12.22  & T Pyx/20.5  & -7.1/-6.8  &     \\
			SMCN 2005-08a     & 3588.1  & -/(10)  & U Sco/22  & -/-8.6(-9)   &     \\
			SMCN 2006-08a     & 3952  & 11.8/-  & T Pyx/21  & -7.2/-  &     \\
			SMCN 2012-09a     & 6185  & 12.1(9.8)/-   & U Sco/20.4   & -6.9(-9.2)/-  &     \\
			\bottomrule
		\end{tabular*}
		\begin{tablenotes}
			\item Note: An estimate of the brightness in a maximum is given in parentheses. ? - under doubts. 
		\end{tablenotes}
	\end{table*}
\end{center}

When constructing the light curve of LMCN 2005-11a (Fig.14), we relied on the OGLE survey data. The OGLE monitoring missed only 3 days of the initial brightness increase and, starting from the final brightness rise, is overlap during 200 days the BVR photometry of \cite{Liller2007}. With the parameters of Table 2, the light curve closely matches the light curve of T Pyx: the amplitude of the outburst, the presence of a prolonged state of maximum brightness, with the exception of a later onset of the final decline. Photometry convincingly showed the presence of brightness variations of LMCN 2005-11a at the maximum state. We did not use the SMARTS data, since the I photometry data after a brightness of I$\approx$15$^{m}$ began step by step to rise above the OGLE data and in the last dates of observations the difference reached 5.5$^{m}$ \cite{Walter2012} noted a similar general problem and worked with it). V-R color index according to \cite{Liller2007} virtually coincides with V-R of \cite{Walter2012} on an overlapping observations interval. A comparison of the OGLE photometry and photometry of \cite{Liller2007} allows us to estimate V-I$\approx$1$^{m}$ over all overlapping time interval. But the V-I index from the SMARTS survey from the first dates of observations was about 4$^{m}$. 

LMCN 2013-10a \cite{Mroz2016b} is classified as a member of the T Pyx group with the parameters of Table 2 (Fig.14). 

The existence of a large group of recurrent novae of the T Pyx type gives us the opportunity to estimate the range of duration of maximum brightness states to be 28-55 days, which is clearly seen from the range of shifts of the light curves at the final phase of the outburst which is equal $\Delta$log(t-t$_{0}$)$\approx$0.5. 

The maximum brightness of LMCN 2012-03a was recorded by several observers \citep{Seach2012, Walter2012, Mroz2016b}. When using this data set, we focused on the V photometry of the SMARTS survey, therefore the data of \cite{Seach2012} were corrected by +1.2$^{m}$, and from the OGLE data for the first effective date we derived the colour index V-I=0.8$^{m}$. The result in Fig.15 is the satisfactory agreement with T CrB. A secondary outburst like the prototype was not recorded. 

The outburst maximum of OGLE-2012-Nova-003, LMCN 2012-10a, \citep{Wyrzykowski2012, Mroz2016b} falls into a 3-day gap in the observations, which does not radically distort the shape of the light curve after 100 or more days. The duration of the outburst was no more than 430 days, that is not typically for classical nova and more typical for recurrent one. The light curve in the I band is characterized by monotony and the absence of specific details on it. With this data set, we got the best agreement with the V745 Sco group (Table 2). In this case, the nova brightness at the time of obtaining the overexposed image will fall within the boundaries of 11-12$^{m}$ as it was estimated by \cite{Wyrzykowski2012}. The outburst amplitude which is nearly 3$^{m}$ higher of the prototype may be due to the favourable orientation of the binary system. 

Photometry of SMCN 2005-08a of the OGLE survey \citep{Mroz2016b} in the I band did not capture the maximum. But we can supplement OGLE photometry with the visual and BVR photometry near the maximum brightness of \cite{Liller2005} and \cite{Pearce2005}, and also by the non-system data of \cite{Neill2005} at the initial brightness decline phase. Then by correcting the non-system observations according to the color index V-R=0.8$^{m}$, and also correcting of the light curve in the I band on +2.5$^{m}$ we can obtain a satisfactory visual light curve and come to a confident conclusion about the similarity with U Sco (Fig.15, Table 2). The total duration of outburst is about 400 days. This is longer than that of U Sco, but the outburst amplitude is also higher (a peculiar example of homothetic transformation of two curves with respect to the point (1,13.1) from field of geometry). The maximum brightness in the I band can be estimated at about 10.6$^{m}$. 

\begin{figure}[t]
	\centerline{\includegraphics[width=78mm]{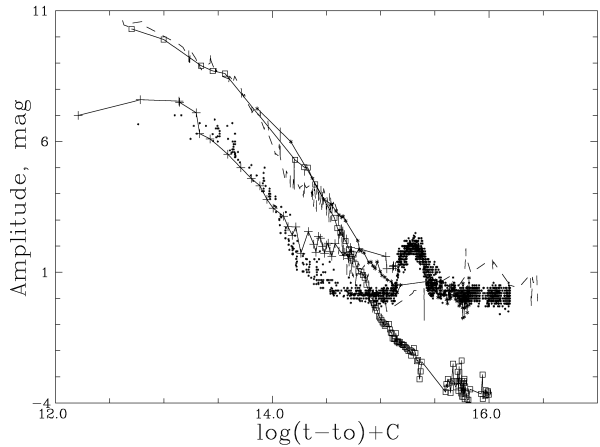}}
	\caption{The light curves of novae in the Magellanic Clouds as candidates in recurrent novae of the U Sco group: U Sco (dashed line), SMCN 2005-08a (line with squares) and SMCN 2012-09a (line with dots), and the T CrB group (left and lower): T CrB (dots), LMCN 2012-03a (left line with pluses). Note that the ordinate scale range is larger than other patterns to display the high outburst amplitude of SMCN 2005-08a.\label{fig15}}
\end{figure} 

Extrapolation of the light curve of SMCN 2012-09a (Fig.15) to the outburst maximum to fully match with the light curve of U Sco is possible with the following parameters. The maximum brightness was probably about 9.8$^{m}$, i.e. the observed value from the OGLE-IV catalog \citep{Mroz2016b} should be increased by 2.3$^{m}$. The maximum time is t$_{0}$=JD 2456185, which corresponds to 1.25 days after the previous OGLE observation, which did not detect the outburst, and the next observation after 2.75 days was already fixed with the outburst (the first point of the light curve of Fig.15). 

Our representation of the light curve of LMCN 2012-03a confirms the assumption of \cite{Schwarz2015} and \cite{Mroz2016b} that this nova is possibly the recurrent nova of the U Sco subclass. But by our opinion the light curve by its shape is closer to T CrB. The smooth light curve of LMCN 2012-03a in the I band has a slope of the light curve similar to the visual light curve of T CrB (Fig.15, Table 2). The final rise of the light curve took a day, then according to visual observations of \cite{Seach2012} a plateau of about 1 day appeared , which was as short as that of T CrB. It is possible that the outburst amplitude was slightly higher than that of T CrB, since \cite{Schwarz2015} discuss in detail the question of 5 close candidates for the post-nova within errors of the measured object coordinates. Close stars of the field may be the reason for the smooth transition of the nova light curve to its quiet state starting from the brightness on 2.5$^{m}$ higher of the level of quiescence brightness. 

LMCN 2010-11a has 5 fixed outbursts. Figure 16 shows the summary of three outbursts. The 2016 outburst \cite{Mroz2016a} is also represented by data in the BVRI bands of \cite{Munari2016}, the 2010 outburst has data in the I band \citep{Mroz2016b}, and for 1968 there are photographic measurements  of \cite{Sievers1970}. Figure 16 shows the satisfactory repeatability of light curves over a time equivalent to 10 recurrent outbursts (recurrence time is estimated to be no more than 5 years by \cite{Munari2016}), similar to 10 of U Sco and 8 of RS Oph. Photographic data of \cite{Sievers1970} and color indices near the outburst maximum \citep{Munari2016, Liller1990} allow us to estimate the V brightness at the maximum of about 10.7$^{m}$. We classified this nova as a possible member of the CI Aql group, since formal matching of light curves is possible if the light curve of LMCN 2010-11a is shifted along the abscissa by $\Delta$log(t-t$_{0}$)=+0.9 (Fig.16). The difference in the development of outbursts is the absence of a plateau at the maximum and only one plateau at the final decline of LMCN 2010-11a, while CI Aql had a prolonged maximum and there were 3 “plateaus” on the final brightness decline. It is possible that such a more rapid development of the LMCN 2010-11a outburst can be attributed to the short, 5-year recurrence time. The matter ejection velocity of these two novae was about 2000 km/s \citep{Shore1991, Iijima2012}). It should be noted that the slope of the initial brightness decline is close to the V745 Sco group.

\begin{figure}[t]
	\centerline{\includegraphics[width=78mm]{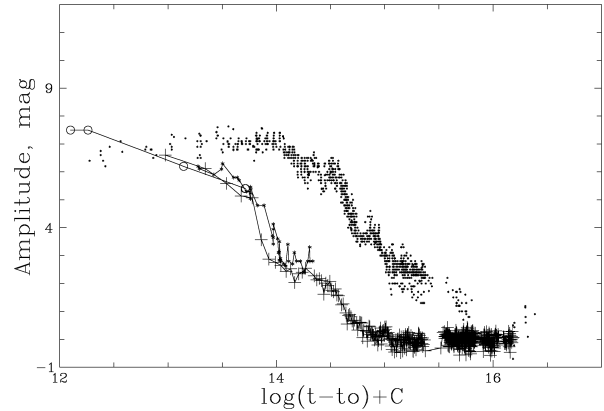}}
	\caption{The light curves of CI Aql (dots) and of a possible member of the CI Aql group – of the recurrent nova of LMCN 2010-11a for outbursts in 1968 - line with circles, 2010 - line with pluses, 2016 - line with dots.\label{fig16}}
\end{figure} 

We included LMCN 2004-10a \citep{Liller2004a, Mroz2014, Mroz2016b, Pietsch2010} in a group with RS Oph (Table 2, Fig.17). It is noteworthy that the outburst amplitude is equal to the amplitude of the candidate for recurrent nova of LMCN 2009-02a (see below) and their light curves differ only in the part of the earlier beginning of the final brightness decline of the first nova.  

We note an interesting situation with the recurrent nova of LMCN 2009-02a \citep{Bode2016} (Fig.17, Table 2), the modified light curve of which has a good similarity with the galactic recurrent nova of RS Oph, but with a outburst amplitude higher on 3-4$^{m}$. The first point is the brightness estimate of V$\approx$11.4$^{m}$ after correction of the observational data without a filter (corresponding to approximately the R band) with the extrapolated color index of V-R$\approx$0.8$^{m}$ \citep{Walter2012} at the outburst maximum. 

Slight differences in the light curves of classified objects from “standards” may arise due to a comparison, for example, of visual observations with data in photometric I band, observations in which form the bulk of the OGLE survey data. 

\begin{figure}[t]
	\centerline{\includegraphics[width=78mm]{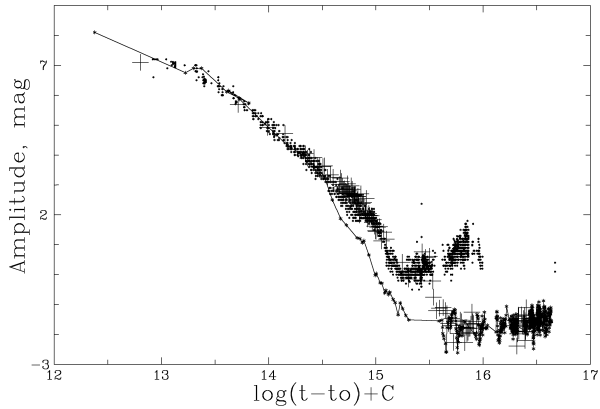}}
	\caption{Light curves of RS Oph (points) and the recurrent novae LMCN 2004-10a (line with dots) and LMCN 2009-02a (pluses) with parameters of Table 2. The zero-point of the amplitude scale is shifted by -3$^{m}$ for convenience displaying.\label{fig17}}
\end{figure} 

From the list of recurrent novae candidates of \cite{Rosenbush2002}, we were not able to decide on LMCN 1991-04a, which \cite{Schwarz2001} define as “extraordinary”. The light curve of LMCN 1991-04a in the maximum has a domed shape with the duration of about 5-8 days, unlike many others. The general trend of a brightness decline has the slope closer to recurrent novae than to classical ones. But among recurrent novae, it was also not possible to find an analogue in the rate of brightness decline. By formally varying the parameter t$_{0}$, one can fit the shape of the light curve to one of the fastest classical nova V5579 Sgr (we will return to this question in the second part of the study).

\subsection{Absolute magnitudes of recurrent novae}
 
Conclusion 3 of \cite{Schaefer2010a} means that at the outburst maxima the absolute magnitude of the recurrent nova is always the same. We will extend this conclusion to the entire group of novae (CI Aql, T Pyx, V745 Sco, and other similar groups), which includes this recurrent nova. Then, having determined the absolute magnitude of one or several novae, we can attribute this absolute magnitude to any member of group. What is especially important, in this case, no additional studies are required, for example, interstellar absorption in this direction, or other assumptions. It can be expected that a similar approach can be applied to classical novae too. 

In fact, the final presentation of the light curve of an erupted star, recurrent or classical nova, can be carried out directly in the absolute magnitudes scale. The first and only step for this is to establish belonging to a particular group of novae. The light curves of the novae – of standards of group -is to be represented in this absolute scale. This opportunity is provided to us by novae of galaxies with known distances. Some galaxy are monitored by a network of automated observation complexes. The problem here is with the coverage of the entire amplitude of the light curve in the outburst for reliable classification. Undoubtedly, the nearest galaxies, such as the Large and Small Magellanic Clouds (MCs) with their diverse studies, are the most promising in this regard. 

The issue of interstellar extinction for the MCs was solved by results studying field stars \citep{Haschke2011}). Fortunately, the absorption in the V band in the vicinity of the novae in the LMC was low: all values were in the range 0.05-0.3$^{m}$. Since these values relate to field stars, for the extinction correction of our objects, we used the average interstellar absorption A$_{I}$$\approx$0.1$^{m}$ for the I band and A$_{V}$$\approx$0.2$^{m}$ for V. In the SMC, the absorption is even less: A$_{V}$$\approx$0.1$^{m}$ and A$_{I}$$\approx$0.05$^{m}$. At this phase of the study, these values can be applied to our objects, since the possible differences will not be high, and the scatter of the absolute brightness values for novae can be significantly higher due to individual differences.

The distance modules to the LMC and the SMC according to one of the latest studies \citep{Graczyk2014}: 18.493$^{m}$ and 18.951$^{m}$, respectively. The error in the distance modulus estimate is at about 0.07$^{m}$, the error of distance - about 5 kpc, or the systematic error - at about $\pm$1.11 kpc \citep{Pietrzynski2013}. 

Observations of the OGLE project are carried out mainly in the I band with rare measurements of the V magnitude. At the same time, the catalogues and fundamental data on celestial objects are oriented mainly to the V band, so we first turn our attention to novae of the MCs with such data. 

There is an opinion of \citep{Bond2004} that LMCN 2004-10a is possible to identify its with YY Dor, the nova of 1937, but coordinates for the latter are no exact. Several observations of 2004 of \citep{Liller2005} give a maximum brightness of V$\approx$10.8$^{m}$ for JD 2453298.693. This corresponds to the absolute magnitude of LMCN 2004-10a M$_{V}$$\approx$-7.9$^{m}$. The B-V color index near the maximum was -0.14$^{m}$ and changed to  0.26$^{m}$ after three days. Consequently, the normal colour index for the maximum could have a value of about (B-V)$_{0}$=(B-V)–A$_{V}$/3.2$\approx$-0.21$^{m}$. Analysis of the light curves of LMCN 2004-10a and V745 Sco convinces us that the available photometry of the first star in the visual and near infrared spectral regions corresponds to the extreme brightness in the outburst, and the same conclusion is true for visual observations of the second nova. 

LMCN 2005-11a has been classified by us as a member of the T Pyx group. Detailed BVR photometry of \cite{Liller2007} gives a maximum brightness of V$_{max}$=11.5$^{m}$, which leads to the absolute magnitude in the V band: M$_{V}$=-7.2$^{m}$. In the first 30 days, the B-V color index changed slightly and was about 0.17$^{m}$ \citep{Liller2007}, i.e. normal color index (B-V)$_{0}$=(B-V)–A$_{V}$/3.2$\approx$0.10$^{m}$. 

In 2016, the fifth outburst of LMCN 2010-11a \citep{Munari2016} occurred, the first was in 1968 \citep{Sievers1970}. Above, we estimated the maximum V magnitude of about 10.7$^{m}$. Thus, we have the absolute brightness M$_{V}$=-8.0$^{m}$. 

For the candidate for recurrent novae LMCN 2012-03a with the data of Table 2, we obtain an absolute brightness estimate of M$_{V}$= 6.8$^{m}$. 

Need visual observations LMCN 2013-10a are not available. Therefore, we can derive absolute brightness only in the I band \citep{Mroz2016b}: M$_{I}$=-7.4m. 

The CCD photometry of SMCN 2001-10a in the next day after the discovery gave V=12.22$^{m}$ \citep{Liller2001}. This corresponds to an absolute magnitude M$_{V}$ of about -6.8$^{m}$. 

The only estimate of the extreme brightness of 10.4$^{m}$ \citep{Liller2005} for SMCN 2005-08a most likely corresponds to the maximum of the visual brightness, therefore, we derive the absolute magnitude M$_{V}$=-8.6$^{m}$. 

SMCN 2012-09a has not the visual photometry, therefore the possible extrapolated value M$_{I}$=-9.2$^{m}$ is additionally indicated in parentheses in Table 2. 

The absolute magnitudes in the V-band from Table 2 of \cite{Shafter2013} are presented in the 6 column of our Table 2. 

Now we compare the absolute magnitudes of the MCs and galactic novae (Table 3). 

\begin{center}
	\begin{table}[t]%
		\centering
		\caption{Absolute M$_{V}$ magnitudes of recurrent novae by groups.\label{tab3}}%
		\tabcolsep=0pt%
		\begin{tabular*}{250pt}{@{\extracolsep\fill}rccc@{\extracolsep\fill}}
			\toprule
			\textbf{Galactic$ $ $ $}& \textbf{M$_{V}$ }& \textbf{Nova of}& \textbf{M$_{V}$} \\
			\textbf{nova $ $ $ $}& \textbf{\citep{Schaefer2010a}}& \textbf{the MCs}& \textbf{given study} \\
			\midrule
			CI Aql   & -7.1$^{m}$  & LMCN 2010-11a  & -8.0$^{m}$ \\
			V2487 Oph  & -7.4   & LMCN 2013-10a   & -7.4 \\
			\midrule
			IM Nor  &  -6.6  & LMCN 2005-11a   & -7.2 \\ 
			T Pyx   & -7.1   & SMCN 2001-10a  & -6.8  \\ 
			\midrule
			 T CrB &	-7.6 & LMCN 2012-03a &	-6.8    \\ 
			 \midrule
			 V745 Sco &	-8.2 &	LMCN 2004-10a &	-7.9 \\
			 \midrule
			  U Sco &-8.5\tnote{$\dagger$} & SMCN 2012-09a & (-6.8)-(-9.2)\tnote{$\dagger$} \\ 
			  &    & SMCN 2005-08a  & -8.6  \\ 
			  \midrule
			 RS Oph	& -10.6 &	LMCN 2009-02a &	-7.3  \\ 
		  	\bottomrule
		\end{tabular*}
	\begin{tablenotes}
		\item[$\dagger$] See text for details. 
	\end{tablenotes} 
	\end{table}
\end{center}

The absolute magnitude of SMCN 2012-09a was defined in the photometric I band and whereas the absolute brightness of U Sco is in the V (Table 3). For a galactic object, the absolute brightness of M$_{I}$ can be estimated using multicolour observations of the outburst in 2010 \citep{Pagnotta2015}. The observed V-I$_{C}$ color index at the peak of the outburst was about 0.6$^{m}$. The excess color for U Sco E(B-V)=0.2$^{m}$ \citep{Schaefer2010a}, therefore, E(V-I)$\approx$0.1$^{m}$ and the normal color index (V-I$_{C}$)$_{0}$$\approx$0.5$^{m}$. Thus, we have the absolute brightness U Sco M$_{I}$=-9.0$^{m}$. The proximity of these two absolute magnitudes, according to \cite{Schaefer2010a} and our definition through the MCs (Table 3), indicates that our extrapolated maximum brightness SMCN 2012-09a in the outburst (Table 2) is close to the actual value. 

A comparison of the columns 2 and 4 of Table 3, shows good agreement, based on the simple indirect method proposed here for estimating the absolute brightness of galactic novae. The biggest difference, at 3.3m, is for RS Oph. For the other 6 recurrent novae differences do not exceed 1$^{m}$. Such a large difference for RS Oph can be associated with the individualities of RS Oph and LMCN 2009-02a: the difference in absolute brightness (3.3$^{m}$) is comparable with the difference in the amplitudes of the outbursts of the galactic nova and the LMC nova ($\approx$2.5$^{m}$) (Fig.17). 

The error $\sigma$ of our approach consists, at least, of errors in the distance modulus of the MCs and the unknown position of the nova due to the geometric thickness of the MCs $\sigma$$_{1}$, error in determining the interstellar extinction $\sigma$$_{2}$ for a particular object, instrumental error in the observations of $\sigma$$_{3}$, and the actual spread of the maximum absolute magnitude of nova $\sigma$$_{4}$: $\sigma$$^{2}$ = $\sigma$$_{1}$$^{2}$ + $\sigma$$_{2}$$^{2}$ + $\sigma$$_{3}$$^{2}$ + $\sigma$$_{4}$$^{2}$. For example, taking into account the geometric thickness of the LMC and the SMC, respectively about 4 kpc and 5 kpc \citep{Graczyk2014, Pietrzynski2013} will reduce the error $\sigma$$_{1}$ by about 0.1-0.2$^{m}$, or increase in the same value the accuracy of estimating the absolute value of the nova. 

\subsection{Candidates for recurrent novae among classical novae}  

Candidates for recurrent novae among the classical novae named many researchers. \cite{Duerbeck1987b, Duerbeck1988} proposed the use of the diagram "outburst amplitude, rate of brightness decline after maximum" (the amplitude - t$_{3}$-time diagram) to search for recurrent novae among classical ones. 

\cite{Pagnotta2014} presented the results of an extensive search for candidates in recurrent novae among classical novae. They developed 7 detailed criteria for a search of recurrent candidates among classical novae and compiled the list of 10 candidates. The first two criteria ((1) outburst amplitude smaller than 14.5-4.5$\times$log(t$_{3}$), (2) orbital period >0.6 days) of 7 developed criteria can be obtained as a result of simple photometric observations, and the first of them is a characteristic of the light curve, on the basis of which the classification scheme proposed in this study is based. The rest require spectral observations, for which resources are not always available. 

We can note that the results of studies in 1999-2002 \citep{Rosenbush1999a, Rosenbush1999b, Rosenbush1999c, Rosenbush1999d, Rosenbush2002} and subsequent experience convinces us of a clear distinction between the shape of the modified light curves of known recurrent novae and classical novae. The difference in the amplitudes and duration of outburst in our scheme is transformed into such a property: the light curve of the recurrent nova in the modified scale of time has a greater slope in the final decline phase. It should be understood that recurrent novae do not have the transition phase, unlike classical novae. Doubtful or erroneous identification, classical or recurrent given nova, is possible only in case of an incomplete light curve. 

The initial selection of candidates for recurrent novae was carried out on the basis of signs of a low outburst amplitude and a short outburst duration. The results are presented in Table 4. Naturally, we were looking for analogues of the known recurrent novae, and there may also be unknown types of recurrent ones. First of all, we will consider simpler versions of novae with characteristic details on the light curves: the same slope of the light curve, the presence of a plateau at the maximum light, stops (steps) during the brightness decline. And we end our search by novae with smooth light curves: linear throughout or smooth curves, which may make the comparison ambiguous. The last column of Table 4 shows the category belonging of the object to the classical or recurrent novae according to the list of \cite{Pagnotta2014}  (i.e., as of 2014). Recall that category A contains the known RN, category B contains strong RN candidates, category C contains likely RN candidates, category D contains likely CNe, category E contains systems which are certainly CNe, category F contains systems for which there is not enough information.

\begin{center}
	\begin{table*}[t]%
		\caption{Candidates for recurrent novae and their membership in the group of recurrent novae or similarity to the particular recurrent nova.\label{tab4}}
		\centering
		\begin{tabular*}{400pt}{@{\extracolsep\fill}rlllrc@{\extracolsep\fill}}
			\toprule
			\textbf{Nova} & \textbf{Year}  & \textbf{t$_{0}$}  & \textbf{A/Accepted V$_{q}$} & \textbf{Prototype}  & \textbf{Category} \\
			\midrule
			V2672 Oph &	2009 &	45058.5  &	7.0/18(20) &	T CrB &	- \\
			V1534 Sco &	2014 &	56742.5	 &  7.5/18.5 &	T CrB & 	- \\
			V1707 Sco &	2019 &	8741.3 &	7/18.8 &	T CrB &	- \\ 
			KT Eri &	2009 &	55149 &	7.0/12.5 &	RS Oph &	- \\
			QZ Aur &	1964 &	38418 &	10.0/16p &	T Pyx &	D \\
			RS Car &	1895 &	13266 &	9.3/16.5p &	T Pyx &	F \\
			BT Mon &	1939 &	29500 &	8.9/16.5p &	T Pyx &	E \\ 
			V553 Oph &	1940 &	29810 &	9/21.5p &	T Pyx &	- \\
			GR Sgr &	1924 &	23790 &	9:/15.5p &	T Pyx &	E \\
			V787 Sgr  &	1937 &	28672 &	10/19p  &	T Pyx &	D \\ 
			V1017 Sgr &	1919 &	21981&	(9.5)/14.5p&	T Pyx&	E \\
			V3964 Sgr&	1975&	42564&	9.5/18&	T Pyx&	F \\  
			V4092 Sgr&	1984&	45966&	9.8/19.5 &	T Pyx&	- \\ 
			V5852 Sgr&	2015&	57084&	9.2/22&	T Pyx&	- \\ 
			V711 Sco&	1906&	17315&	9.8/19.5&	T Pyx&	-\\ 
			V556 Ser&	2013&	56610&	9.3/21&	T Pyx&	- \\ 
			V1705 Sco	&2008&	54470&	8.5/21&	T Pyx&	- \\ 
			V5861 Sgr&	2010&	55370&	9/22 &	T Pyx&	- \\
			V1428 Cen&	2012&	56022 &	7.9/17.5 &	-&	- \\ 
			V4739 Sgr &	2001&	52147.5&	9.6/17.5&	U Sco&	B \\ 
			V1311 Sco &	2010	&55311&	10.7/19&	U Sco&	- \\ 
			V1533 Sco	&2013	&56441.5&	10.5/21&	U Sco&	- \\ 
			HV Cet&	2008&	54605&	(7)/18&	CI Aql&	- \\ 
			LS And &	1971&	41187&	8.8/21.5&	V745 Sco&	F \\
			DE Cir&	2003&	52921.35&	9.6/17.3 &	V745 Sco&	B 	 \\
			\bottomrule
		\end{tabular*}
		\begin{tablenotes}
			\item Note: p - photographic magnitude.  
		\end{tablenotes}
	\end{table*}
\end{center}

The photometric and spectral properties of Nova Oph 2009, V2672 Oph, led to a discussion of \cite{Munari2011} about the recurrent nature of this fast nova. Of the three variants of the similarity of T CrB, RS Oph, or U Sco, they settled on the latter: both novae had similar plateaus in the middle part of the brightness decline, similar profiles of extremely wide emission lines. The recurrent nature of V2672 Oph is also confirmed by the light curve in the modified scales (Table 4). With RS Oph, the V2672 Oph light curve does not have any similarities. Rather, it is intermediate between T CrB and U Sco. [For a better comparison of light curves, we recommend shifting the light curve of V2672 Oph along the abscissa to +0.1, which is permissible and can be attributed to errors in the procedure (see, Section 2 with the procedure).] V2672 Oph differs from U Sco in the prolonged plateau in the middle of the light curve, but has an equal outburst amplitude if the maximum brightness is extrapolated to the moment JD 2455059$\pm$0.5 and the peak brightness to the magnitude V=7.0$^{m}$. (The light curve of \cite{Munari2011} ends at the very beginning of this plateau, that well traced by the AAVSO and SMARTS data \citep{Walter2012}.) The T CrB plateau before the secondary outburst is comparable to V2672 Oph, but the plateau of the second nova ended the further decline in the brightness. We include V2672 Oph in the T CrB group. 

Nova Scorpii of 2014, V1534 Sco, has already been proposed to be considered as a recurrent nova similar to T CrB based on the similarity of physical characteristics \citep{Joshi2015}. Similarly to \cite{Joshi2015} we obtained the light curve of V1534 Sco from two sources \citep{Kafka2019, Walter2012} with the parameters of Table 4. Here can note the possible secondary increase in V1534 Sco brightness by analogy with T CrB. 

V1707 Sco very well coincides with the light curve of the prototype of T CrB, it remains to wait for a secondary brightening from January 2020 from 18.8$^{m}$ to 16.8$^{m}$ in February-March 2020 and the subsequent final return to the quiescence state. Our classification is in conflict with the spectral one by \cite{Strader2019} that determined V1707Sco as a classical nova. 

The KT Eri classification is interesting due to the wide discussion about the possible belonging to recurrent novae (see, for example, \cite{Jurdana-Sepic2012}), as indicated by some characteristics: the high expansion velocity, development of the X-ray radiation. The AAVSO data were supplemented by publications of \cite{Bruch2018, Hounsell2016}, the post-nova brightness was assumed to be 14.8$^{m}$ \citep{Shafter2013}. The light curve of this nova went through a comparison procedure with all the groups of recurrent and classical novae that we identified. The choice between the belonging to the RS Oph group or less confidently to any group of classical novae, CP Pup or Her1, was made in favor of the former. To match the light curves in both cases, it is necessary to equally increase or decrease the brightness of the quiescence state: by 2.2$^{m}$. In the case of the RS Oph group, the coincidence of the light curves of the three novae, RS Oph itself, LMCN 2009-02a and KT Eri, is more obvious: from the maximum brightness to the abscissa 15.1 they are almost identical, further follows a small short-term depression and the RS Oph brightness reaches a level of the quiescence, but the brightness of KT Eri and LMCN 2009-02a continued to decline still on 2$^{m}$ (Fig.17). The outburst duration of about 600 days of KT Eri is large than nearly 150 days for RS Oph but less than for the classical novae of V446 Her and CP Pup: about 1500 and 2400 days correspondingly. The coincidence of the light curves of KT Eri and LMCN 2009-02a was for us the last argument for the recurrent nature of KT Eri.

We ascribed the old nova of QZ Aur \citep{Gessner1975} to the T Pyx group (Table 4). In this case, we attribute the beginning of the state of maximum brightness to date JD 2438418. The difference between our estimate of the brightness of the quiet state (m$_{pg}$=16$^{m}$) and modern observations of \cite{Campbell1995} (B=17.50$^{m}$) or the average brightness (B=17.1$^{m}$) before the 1964 outburst \citep{Schaefer2019} can be associated with the spatial orientation of the binary system or a physical variability.

The light curve of RS Car, which burst in 1895 \citep{Walker1933}, allows us to propose this nova as a candidate for the recurrent one. The only resemblance to the classical CP Pup group belongs to the non-typical short depression, but which is following by the atypically early start of the final decline in brightness. With the parameters of Table 4, the light curve satisfactorily coincides with the light curve of T Pyx with the exception of a local re-brightening (or the short depression, see above) on 1.5$^{m}$ near log(t-t$_{0}$)$\approx$15.1  (JD 2413360). In this interpretation, the data of \cite{Walker1933} recorded the maximum brightness.

Our classification of the outburst of Nova Monocerotis 1939, BT Mon \citep{Whipple1940, Bertiau1954}, as the recurrent nova of the T Pyx group, assumes that its outburst began on a date between JD 2429500 and JD 2429515 (the last date is according to \cite{Payne-Gaposchkin1957}).

V553 Oph has a small amount of observational material (8 photographic plates) \citep{Burwell1941}, but its interpretation in our scheme has a small selection (Table 4). The prolonged state of maximum brightness (about 20$^{d}$), the fast initial decline in brightness and the outburst amplitude form the light curve typical to the T Pyx group. \cite{Mroz2015} provide the following data for a possible candidate for the old nova of V553 Oph: the distance from catalog coordinates is 0.75", the brightness V=21.382$^{m}$ and the color index V-I=2.552$^{m}$. This post-nova brightness is close to our estimate.

The GR Sgr outburst was already detected at the final phase \citep{Woods1927}. The duration of the outburst was less than 740 days, which is not typical of the classical nova. The only way to classify the light curve of this nova is to define it as a candidate for the T Pyx group (Table 4). Extrapolating the maximum gave some additional parameters: the final brightness rise began in JD 2423790 from the photographic brightness of about 7.5$^{m}$ and near JD 2423820(-3/+5) the nova reached the maximal value of about 6.5$^{m}$, which is only 1$^{m}$ brighter than the \cite{Dishong1955} estimate: mpg$\approx$7.5$^{m}$. \cite{Mroz2015} give the brightness V=16.298$^{m}$, V-I=0.531$^{m}$ for a possible candidate for a post-nova at a distance of 0.03 arcsecond from the coordinates of the SIMBAD database, which is close to our estimate of the brightness of a quiescent (Table 4).  

V787 Sgr, despite the incomplete light curve \citep{Swope1940}, can be considered as a candidate for the recurrent nova of the T Pyx groups, given the slope of the initial light decline and the photographic magnitude of the quiescence state. Our estimate of this parameter (Table 4) agrees well with the observational data of a possible candidate of \cite{Mroz2015}: V=19.122$^{m}$, V-I=1.343$^{m}$. 

An interesting example of applying our classification is with V1017 Sgr. This nova over more than 100 years of observations had several outbursts of different amplitude and duration \citep{Salazar2017}. Outbursts of the dwarf nova type in 1901, 1973 and 1991 lasting several months had amplitudes of about 3m and they are beyond the scope of our study. The 1919 outburst with a duration of about a year and an amplitude of more than 7.2$^{m}$ was classified as a classical nova outburst, and this is of interest to our study. \cite{Salazar2017} provide a critical review of all available photometry. Applying our methodology to the outburst in 1919, we came to the conclusion that this was an outburst of a typical recurrent nova of the T Pyx group (Table 4). This gave us the opportunity to reconstruct the main parameters of outburst: the beginning of the outburst was near JD 2421981, the duration of the maximum brightness state was about of 40-45 days, the maximum brightness m$_{pg}$=5$^{m}$ was achieved near JD 2422000. This interpretation of the 1919 outburst leads to a photographic brightness of the quiet state of 14.5$^{m}$, which coincides with the data of \cite{Salazar2017}: B$\approx$14.7$^{m}$. I.e. V1017 Sgr has outbursts both a recurrent nova of the T Pyx type and a dwarf nova type.  

The light curve of the V711 Sco \citep{Walker1933} is identical to T Pyx (Table 4).

The light curve of the V556 Ser \citep{Itagaki2013, Kafka2019} with the parameters of Table 4 is in good agreement with the prototype T Pyx. The peak brightness at 13$^{m}$ the nova reached near JD 2456610. 

V1428 Cen has a light curve \citep{Kafka2019, Walter2012} unusual for a classical nova. By sign of \cite{Duerbeck1987b, Duerbeck1988} it can be a recurrent, but peculiar nova, since it is not similar to any known recurrent nova (Table 4). In turn, according to IR photometry of \cite{Walter2012}, there was the small IR excess, which is not typical for recurrent novae. 

To bring the light curve of V1311 Sco \citep{Nishiyama2010a} into coincidence with U Sco, it was only necessary to lower the actual amplitude of the former by 2$^{m}$, which is shown in Table 4 as a higher brightness of the quiet state in comparison with the data of \cite{Walter2012}. 

V1533 Sco has a good set of observational data \citep{Nishiyama2013, Kafka2019, Walter2012, Mroz2015}, which allows us to confidently classify it as a recurrent nova with the U Sco prototype. The combination of the visual light curve \citep{Kafka2019} and photometry of \cite{Walter2012} resulted in a light curve long enough in the scale of the light amplitude and the logarithm of time to perform a subsequent comparison with the light curve of the OGLE \citep{Mroz2015}. As a result, we can confirm our above conclusion that the light curves in the I band have slightly  more negative slope than the visual light curves. 

\cite{Pejcha2008} reported discovered by the Catalina Real-time Transient Survey the outburst of HV Cet and on the similarity of the observed spectral features with spectra of recurrent nova CI Aql, assigned to a nova-like phenomenon. Just later \cite{Prieto2008} given arguments in favour of an O/Ne nova and given also a photometry for pre- and post-season of conjunction of object with the Sun. With data of \cite{Pejcha2008, Zemko2018, Walter2012} we can plot the light curve of HV Cet with a quiet state before season of invisibility and trace the brightness decline up to 2018. The modified light curve of HV Cet with the parameters of Table 4 well coincides with CI Aql. Actual amplitude of outburst, apparently, on 1.2$^{m}$ above, than at CI Aql, but it is in permissible limits of deviations because of space orientation relatively to the observer \citep{Warner1987}. The maximum visual brightness of about 11$^{m}$ was near JD 2454605. In the second part of our study, we will return in assessing whether this nova belongs to the given group of recurrent novae: we weaken this possibility. 

The last group in our consideration of candidates for recurrent novae is the V745 Sco group. The group’s summary light curve up to log(t-t$_{0}$)$\approx$14.0 has the characteristic slope of the initial linear brightness decline, the further brightness decline occurs within a band with the width of $\Delta$log(t-t$_{0}$)$\approx$0.3. The little-studied nova LS And \citep{vandenBergh1973} has the light curve with a length that is typical to a recurrent nova and can be included in the V745 Sco group by the shape of the light curve (Table 4). \cite{Duerbeck1988} and \cite{Pagnotta2014} also included LS And in the candidates for recurrent novae too. A member of this group may also be DE Cir.  

For candidates in recurrent novae of our preliminary list \citep{Rosenbush2002}, we supplemented the observational data from the compilation of \cite{Strope2010} and the AAVSO database \citep{Kafka2019}, which made it possible to clarify the preliminary qualifications. We will postpone the solution of the issue with two novae, V606 Aql and V630 Sgr, until we consider classical novae. Similarly, two novae, V697 Sco and V373 Sct, were excluded from the list of candidates for recurrent ones and we will also return to them in the second part of this publication. The remaining two novae, V3964 Sgr and V4092 Sgr, remain in the list of candidates (Table 4). Their light curves have a maximum brightness state of up to 10 days and the slope of the light curves at the initial decline phase of light curve is typical of the T Pyx group. The brightness of the first post-nova is given in the study of old novae of \cite{Tappert2016}: R=19.0$^{m}$. The maximum state according to \cite{Goranskij1978} is shorter than 10 days, since an increase in its duration leads to an atypical increase in the negative slope of the final brightness decline. The light curve of V4092 Sgr \citep{Liller1984, McNaught1984} must be additionally shifted to the left along the abscissa axis to combine with the prototype, which may be due to the short duration of the maximum state. Combination with T CrB requires both a similar shift in abscissa and in the amplitude scale (by about (-2)-(-3)$^{m}$). The OGLE survey data \citep{Mroz2015} gives the brightness of a possible candidate for V4092 Sgr: V=20.970$^{m}$, V-I=2.067$^{m}$. Both values, V3964 Sgr and V4092 Sgr, do not contradict our estimates of a quiet state brightness (Table 4). 

Now we also pay attention to a number of novae that were discovered during the OGLE survey \citep{Mroz2015} and have observations only in the I band, which may raise reasonable doubts about the reliability of our classification, but additional attention should be to these objects in the future. 

The light curve of V5852 Sgr (OGLE-2015-Nova-01) \citep{Aydi2016}, according to the OGLE survey, represents a outburst from the beginning to the end, and by its shape, this nova can be included in our list of candidates for recurrent novae of the T Pyx groups (Table 4). The increased coefficient of slope of the light curve in comparison with the prototype can be associated with the difference in the photometric bands: I-band in the first and visual for the prototype. A positive argument is the stopping in the middle of the final brightness decline near the 4$^{m}$ level.  

The outburst duration of OGLE-2010-Nova-01 (V5861 Sgr) \citep{Mroz2015} definitely indicates that this nova is recurrent one and with the parameters of Table 4, the light curve fits well with the prototype of T Pyx. A limited observation gap of 37 days makes it possible to optimally select the date t$_{0}$=JD 2455370 for the nova enter in the state of maximum brightness lasting no more than 7 days, which forms the desired slope of the light curve at the phase of returning to a quiet state. 

OGLE-2008-Nova-01 (V1705 Sco) was observed only in the framework of the OGLE survey \citep{Mroz2015} and observations began after the invisibility season, but covered most of the light curve (Table 4). This made it possible to confidently include OGLE-2008-Nova-01 in the candidates for the T Pyx group recurrent novae. According to our estimation, the state of maximum brightness lasted about 35 days.

A candidate in a recurrent nova is V5586 Sgr, which has an atypical light curve for a classical nova with an amplitude more typical for a recurrent one \citep{Walter2012, Nishiyama2010b}. It is problematic to include this candidate in a certain group of recurrent novae, since the visual light curve has a large scatter of points due to the high color index: V-R$_{C}$$\approx$2.3$^{m}$ \citep{Maehara2010}, which makes it difficult to obtain even an average light curve when data sources are heterogeneous. 

All of these stars need further research. This primarily refers to the detection of a second outburst in the study of photo archives. For example, BT Mon may have a circumstellar shell like T Pyx (see discussion in the study of \cite{Tomov2015}), in the group of which we included this candidate. Somewhat earlier \cite{Tomov2008} spoke about the belonging of V2491 Cyg to recurrent novae, but we did not find any convincing signs for classifying it as recurrent nova (when discussing classical novae, we will return to this nova as a member of a small group of novae with unique light curves). V4739 Sgr \citep{Livingston2001} (Table 4) was already mentioned as possibly recurrent nova in the study of \cite{Pagnotta2014}, we only clarified its belonging to a specific group - U Sco. For other candidates for recurrent novae from the list of \cite{Pagnotta2014}: CP Cru, V4160 Sgr, and V477 Sct, we did not confirm this possibility and clarified they belonging to the certain group of classical novae. 

V2487 Oph is included in the Intermediate Polar Catalog of \cite{Mukai2014} as a possible intermediate polar. 

\section{Conclusion}\label{sec5} 

This study develops the preliminary conclusion of \cite{Rosenbush1999a,Rosenbush1999b} on the existence of several groups of classical novae with similar light curves, and therefore similar physical parameters of processes in binary systems, which are novae, and unifies certain classification rules. 

Taking advantage of the modified light curves of nova outbursts for a compact representation of the light curve, both in the region of the maximum light and in the final phase of outbursts, we performed a comparative review of the available photometric data for known novae. Modification of the abscissa scale refers to the possibility of transforming it into an actual logarithmic scale of the radius of the shell, which is ejected during a nova outburst. For a unambiguous construction of the modified light curve, a single rule is introduced for determining the moment of outburst maximum. Modification of the ordinate scale - the current brightness scale of a nova - comes down to the proposal to use the outburst amplitude scale, i.e. the brightness of a nova relatively a quiet state that more really reflects the energy realising during of outburst. 

The third outburst of V3890 Sgr with a very detailed light curve from the maximum to the return of the star to a quiet state was an excellent test of the proposed methodology for constructing modified light curves. Here, the important parameter t${_0}$ is known directly from observations to the nearest hundredths of a day and this did not require “subjective” correction of this parameter for the first two outbursts, which means that they were estimated with good accuracy and this adds confidence in the application of the described technique to construct modified light curves of novae outbursts. 

In this part I of our review, the technique for constructing the modified light curve was applied to 10 known recurrent novae. Three important results were obtained. First, in a series of recurrent outbursts, the light curve retained the shape for all novae. In modified scales, this conservation is displayed more clearly and expressively. Therefore, summing up all known outbursts, we can derive a summarized light curve of this nova. Over the course of 150 years, 10 outbursts of U Sco does not give reasons to detect significant differences in the main details of the light curves. This confirms the conclusion and creation of the templates of light curves of known recurrent novae \cite{Schaefer2010a}. The second conclusion follows from this: the outburst amplitude of the recurrent nova hardly changes, or if it changes, then is small. Thirdly, the same shape of the light curve can occur in two and, possibly, more number novae. This confirms our preliminary conclusion about the possible existence of recurrent novae groups \citep{Rosenbush2002}. As a result, 3 groups were distinguished: CI Aql (+ V2487 Oph), T Pyx (+ IM Nor) and V745 Sco (+ V394 CrA, V3890 Sgr). Moving toward lower amplitudes of outbursts in the framework of this study is not planned. This also applies to dwarf novae with recurrent outbursts and possibly with noticeable differences in the shape of light curves in the sequence of these outbursts. 

Three recurrent novae are beyond these groups: T CrB, RS Oph, and U Sco. Each nova has well presented light curve with own peculiarities and forms its own group from candidates in recurrent novae. 

If we agree that at the peaks of outbursts the absolute magnitude of the recurrent nova is always the same and it is the same for all members of the group, then, having determined the absolute magnitude of one or several novae in the group, we can attribute this absolute magnitude to any member of group. What is especially important, in this case, no additional studies are required, for example, for the interstellar absorption in this direction, or other assumptions. It can be expected that a similar approach can be applied to classical novae too. 

Having obtained a set of light curves of the recurrent novae of our Galaxy and comparing with it light curves of novae of the Large and Small Magellanic Clouds, in addition to the 3 known recurrent novae, we selected several candidates for recurrent novae. Their belonging to one or another group of galactic recurrent novae was specified, and, consequently, the physical characteristics of these novae are similar. Distance and interstellar extinction for the MCs are well known. Assigning the absolute magnitude of the corresponding nova of the MOs to galactic novae, we obtained good agreement with the known estimates of the absolute magnitudes of the former \citep{Schaefer2010a}. Our estimates of the absolute magnitude of galactical novae are as follows: CI Aql, V2847 Oph - M$_{V}$=-7.5$^{m}$; IM Nor, T Pyx - M$_{V}$= 7.2$^{m}$; V3890 Sgr - M$_{V}$=-7.9$^{m}$, U Sco - M$_{I}$= 9.0$^{m}$. These estimates of the absolute magnitude of galactic recurrent novae differ from the published data within the limits of the errors of the latter. The application of this classification scheme to novae in the M31 galaxy and other galaxies is promising, but the classification accuracy may be limited by the impossibility of obtaining a light curve at the final phase of the outburst. 

Another result of our research was a list of 26 candidates in recurrent novae among classical novae and, for example, three of them (KT Eri, V2672 Oph, and V4739 Sgr) coincide with a list of 10 recurrent novae candidates of \cite{Pagnotta2014}. [We will turn to the V838 Her from the list of \cite{Pagnotta2014} in the second part of our study.] In this case, some of the classical novae were classified by us as candidates for dwarf novae, which we left outside the scope of our study, since for short outbursts dwarf novae need more detailed light curves. Thus, taking into account the confirmation of \cite{Pagnotta2009} of our assumption \citep{Rosenbush2002} about the recurrent nature of the V2487 Oph, the result of this study was a list of 22 candidates for recurrent novae. There are several novae that can be considered as candidates for recurrent novae, but for them we did not find analogues among known recurrent novae; we will return to them in the second part of our study. Applying our methodology to the V1017 Sgr outburst in 1919 led us to the conclusion that it was the T Pyx type outburst, i.e. V1017 Sgr, along with outbursts of the recurrent nova often shows dwarf nova outbursts. 

The inclusion in one group of recurrent nova candidates with high amplitudes of outbursts, as, for example, SMCN 2005-08a, may mean a difference in the absolute brightness of the quiescent state of binary systems in which these novae broke out. 

We consider this first part of our study as one of the main arguments for constructing a new scheme for the classification classical novae. The second argument will be presented in the second part.


\section*{Acknowledgements}

We thank the AAVSO observers who made the observations on which this project is based, the AAVSO staff who archived them and made them publicly available. The BAAVSS database is acknowledged as the (part) source of data on which this article was based. This research has made use of the AFOEV database, operated at CDS, France. This research has made use of the NASA's Astrophysics Data System and the SIMBAD database, operated at CDS, Strasbourg, France. This work has made use of data from the European Space Agency (ESA) mission Gaia (https://www.cosmos.esa.int/gaia), processed by the Gaia Data Processing and Analysis Consortium (DPAC, https://www.cosmos.esa.int/web/gaia/dpac/consortium) and the Photometric Science Alerts Team (http://gsaweb.ast.cam.ac.uk/alerts). Funding for the DPAC has been provided by national institutions, in particular the institutions participating in the Gaia Multilateral Agreement. At the most early phase of our study the state of novae was traced by the VSNET data and we are grateful all observers for this possibility. The author is grateful to the administration of the MAO NAS of Ukraine for the opportunity to maintain my staff status without any mutual financial obligations. Author is thankful L.Laurits for useful discussion, and to the Valga vallavalitsus, Estonia, for the financial support, which allowed us to carry out this interesting investigation.









\nocite{*}
\bibliography{Rosenbush1}%



\end{document}